\documentclass[11pt]{article}

\usepackage[final]{acl}

\usepackage{times}
\usepackage{latexsym}

\usepackage[T1]{fontenc}
\usepackage[utf8]{inputenc}

\usepackage{microtype}

\usepackage{inconsolata}

\usepackage{graphicx}
\usepackage{titletoc}

\usepackage{times}
\usepackage{latexsym}
\usepackage{hyperref}
\usepackage{url}
\usepackage{microtype}
\usepackage{amsmath}
\usepackage{inconsolata}
\usepackage{graphicx}
\usepackage{multirow}
\usepackage{booktabs}
\usepackage{algorithm}
\usepackage{algpseudocode}
\usepackage{tcolorbox}
\usepackage{wrapfig}
\usepackage{subfig}
\usepackage{times}
\usepackage{amssymb}
\usepackage{placeins}
\usepackage{pifont}
\usepackage[T1]{fontenc}
\usepackage[utf8]{inputenc}
\usepackage{microtype}
\usepackage{inconsolata}
\usepackage{graphicx}

\usepackage{geometry}
\usepackage{array}
\usepackage{caption}
\usepackage{ulem}  % for \underline in tables
\usepackage{hyperref}    % 可选：改进标题格式
\usepackage{xcolor} % 推荐用 xcolor，功能更全
\usepackage{adjustbox} 
\title{GRAPHIA: Harnessing Social Graph Data to \\Enhance LLM-Based Social Simulation}

\author{
{\bf Jiarui Ji$^1$}, {\bf Zehua Zhang$^{2}$},  {\bf Zhewei Wei}$^1\footnotemark[1]$, \\
{\bf Bin Tong$^2$}, {\bf Guan Wang$^2$}, {\bf Bo Zheng}$^{2}$\footnotemark[1]\\
$^1$Gaoling School of Artificial Intelligence,  Renmin University of China, Beijing, China\\
$^2$Alimama Tech, Taobao \& Tmall Group of Alibaba\\
 \texttt{\{jijiarui, zhewei\}@ruc.edu.cn} \\
 \texttt{\{yuzheng.zzh, tongbin.tb, shangfeng.wg, bozheng\}@alibaba-inc.com}
}

\begin{document}
\maketitle
\renewcommand{\thefootnote}{\fnsymbol{footnote}}
\footnotetext[1]{\quad Zhewei Wei and Bo Zheng are the corresponding authors.}
\footnotetext[2]{\quad The work was partially done at Gaoling School of Artificial Intelligence, Beijing Key Laboratory of Research on Large Models and Intelligent Governance, Engineering Research Center of Next-Generation Intelligent Search and Recommendation, MOE, and Pazhou Laboratory (Huangpu), Guangzhou, Guangdong 510555, China.}

\begin{abstract}

% This gap stems from the absence of mechanisms that incorporate global structural feedback into individual decision-making.

% Large language models (LLMs) show promise in simulating human-like social behaviors, but current text-only training paradigms fail to align agent-level interactions with emergent network structures. Social graphs offer high-quality labels that encode both local interactions and global network, yet remain underutilized as supervision for LLMs.
% To sum up, existing approaches suffer from two critical limitations: (i) the absence of a generalizable training framework that leverages graph-structured data to couple micro-level interactions with macro-level social network dynamics, and (ii) the lack of unified, quantitative metrics to assess the divergence between simulated and real-world social networks across datasets.

% , yet text-only training fails to couple microscopic interactions with macroscopic network structures 
% We propose a unified evaluation paradigm for social graph simulation, comprising two complementary settings: Transductive Dynamic Graph Generation (TDGG), a micro-level task focused on node-wise interaction alignment; and Inductive Dynamic Graph Generation (IDGG), a macro-level task requiring reproducing network properties. 

Large language models (LLMs) have shown promise in simulating human-like social behaviors. Social graphs provide high-quality supervision signals that encode both local interactions and global network structure, yet they remain underutilized for LLM training. To address this gap, we propose Graphia, the first general LLM-based social graph simulation framework that leverages graph data as supervision for LLM post-training via reinforcement learning. 
With GNN-based structural rewards, Graphia trains specialized agents to predict whom to interact with (destination selection) and how to interact (edge generation), followed by designed graph generation pipelines.
We evaluate Graphia under two settings: Transductive Dynamic Graph Generation (TDGG), a micro-level task with our proposed node-wise interaction alignment metrics; and Inductive Dynamic Graph Generation (IDGG), a macro-level task with our proposed metrics for aligning emergent network properties.
On three real-world networks, Graphia improves micro-level alignment by 6.1\% in the composite destination selection score, 12\% in edge classification accuracy, and 27.9\% in edge content BERTScore over the strongest baseline. For macro-level alignment, it achieves 35.98\% higher structural similarity and 28.71\% better replication of social phenomena such as power laws and echo chambers.
Our results show that social graphs can serve as high-quality supervision signals for LLM post-training, closing the gap between agent behaviors and network dynamics for LLM-based simulation. Code is available at \url{https://github.com/Ji-Cather/Graphia.git}.

% With GNN-based structural rewards, Graphia trains specialized agents to predict whom to interact with (destination selection) and how to interact (edge generation), thereby coupling agent actions with global network coherence.
% We evaluate Graphia under two settings: Transductive Dynamic Graph Generation (TDGG), a micro-level task focused on node-wise interaction alignment; and Inductive Dynamic Graph Generation (IDGG), a macro-level task requiring reproduction of social networks.
% Our results demonstrate that real-world graphs can serve as high-quality structural supervision for LLM post-training, closing the gap between text-driven tuning and structure-aware behavior. 

% We propose Graphia, a reinforcement learning framework for LLM-based social graph generator that uses real-world graph structures as implicit supervision signals. 

% Our results show that integrating structural feedback into LLM training aligns local agent behaviors with global network dynamics, enabling more realistic and analytically useful social graph simulators. which increases the composite destination selection score by 6.1\%
\end{abstract}

% These results demonstrate that integrating graph structure into LLM training can effectively align local agent behavior with global network dynamics, paving the way toward more realistic and analytically useful social graph simulators. 

\section{Introduction}
\label{sec:introduction}

Social simulation with LLM-based agents has emerged as a powerful paradigm in computational social science~\cite{S3,llm_simulation_survey1}, enabling large-scale exploration of emergent social phenomena such as echo chambers and influence propagation~\cite{piao2025agentsociety,coling_echo_chamber}. These macroscopic phenomena arise from microscopic text-based interactions between LLM-based agents. Consequently, realistic social simulation requires modeling the microscopic actions that drive graph evolution~\cite{mcpherson2001birds}.

Despite this intrinsic micro-macro link, current social graph generators often decouple these two levels. At the macro level, deep-learning based generators focus on structural evolution using node IDs but ignore the semantic text that catalyzes edge formation~\cite{TIGGER, DGGen}; while LLM-based generators rely either on qualitative case studies for evaluation~\cite{piao2025agentsociety,coling_echo_chamber} or context-specific pipelines such as Twitter simulation~\cite{GAG} that lack generalization.
At the micro level, prior work focuses on fine-grained behavioral realism. \citet{zhousotopia} propose SOTOPIA to evaluate how agents interact, while \citet{zhou2025personaeval} focus on predicting who interacts next. Yet these LLM-based generators often overlook the macroscopic evolution of the social graph structure.

Collectively, these approaches face three critical limitations: 
\textbf{(1) Representational Deficiency}: traditional deep-learning based models are confined to modeling graph topology and are incapable of capturing the underlying text-driven activity; 
\textbf{(2) Methodological Gap}: the lack of a generalizable training framework that optimizes micro-level interactions and macro-level structure using social graph data as supervision; and 
\textbf{(3) Evaluation Gap}: the lack of unified metrics to quantitatively measure how well simulated social graphs match real ones in both interactions and structure.

\begin{figure*}[t]
  \includegraphics[width=\linewidth]{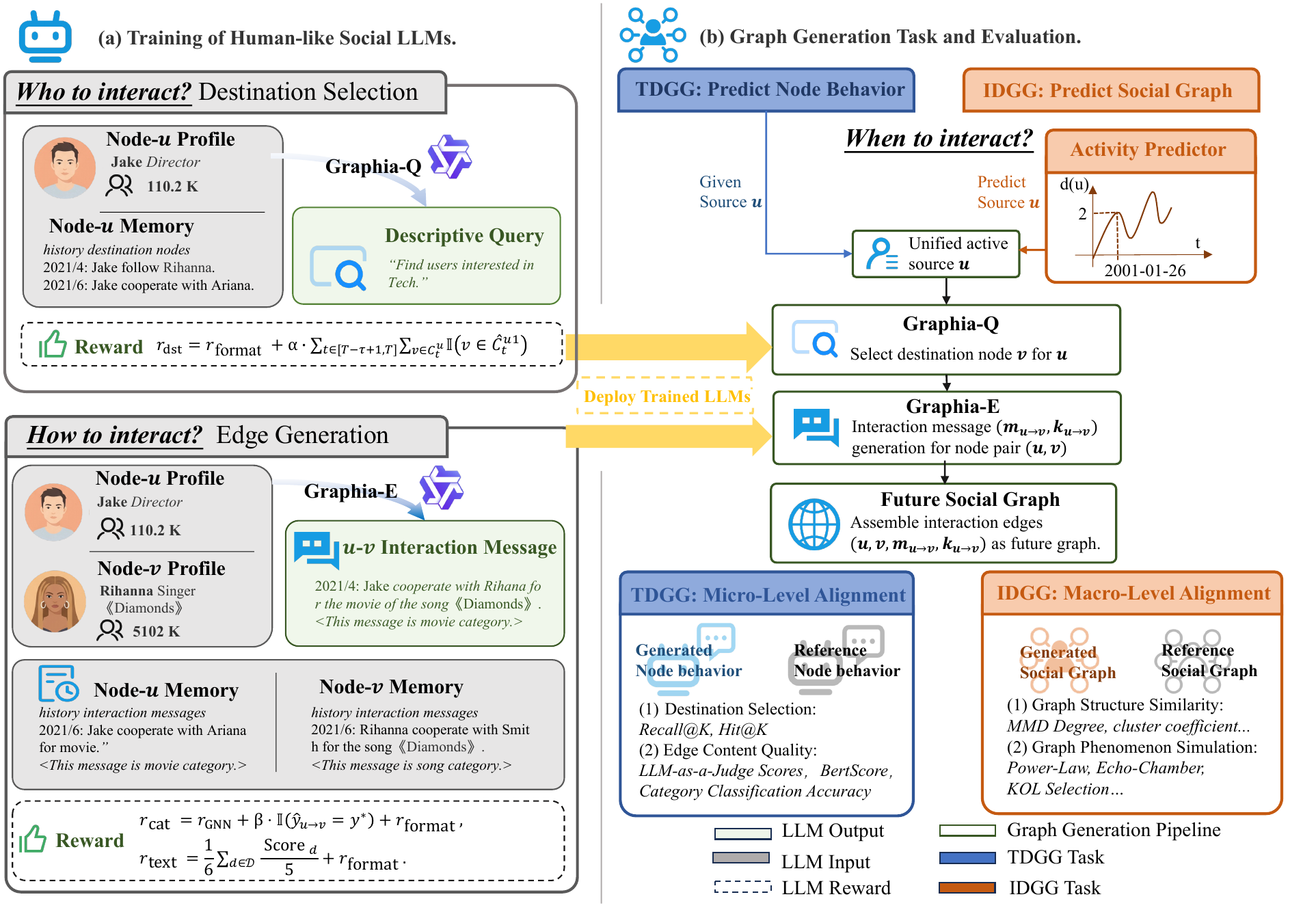}
  % \vspace{1em}
\caption{Graphia training, generation, and evaluation pipeline illustrated on a collaboration network. 
(a) The left panel details the training mechanisms for specialized LLM-based agents: \textbf{Graphia-Q for destination selection} (top-left) and \textbf{Graphia-E for edge generation} (bottom-left). 
These agents leverage text-rich node profiles and interaction memories, with rewards designed to optimize respective tasks. 
(b) The right panel outlines the graph generation pipeline based on trained LLM-based agents for TDGG and IDGG tasks. 
TDGG focuses on micro node behavior; while IDGG, supported by an activity predictor, models the macro social graph.}
\label{fig:framework}
% \vspace{-2ex}
\end{figure*}

To address these limitations, we build on the Transductive (TDGG) and Inductive Dynamic Graph Generation (IDGG) settings from GDGB~\cite{peng2025gdgb} as a foundation for systematic evaluation. In TDGG, we focus on microscopic alignment metrics that evaluate agent-level interactions. In IDGG, we focus on macroscopic alignment metrics that assess whether simulated graphs reproduce real-world graph properties.
Under this paradigm, we formalize three core capabilities for realistic LLM-based social graph simulation:
\textbf{(1) Destination Selection}: Given a source node, can the LLM predict its next interaction partner?
\textbf{(2) Edge Generation}: Can the LLM generate socially coherent and contextually grounded micro-interactions between nodes?
\textbf{(3) Global Structure Fidelity}: Does the generated graph reproduce key macro properties of real graphs?

Guided by these three principles, we propose \textbf{Graphia}, a reinforcement learning framework for LLM-based social graph simulation. Our contributions are: 
(1) The first unified training framework that leverages social graph data as supervision to enhance LLM-based simulation;
(2) A micro–macro evaluation paradigm that extends TDGG and IDGG with novel quantitative metrics for joint assessment of interaction fidelity and network realism;
(3) improved micro-level performance in TDGG tasks, with 6.1\% gain in the composite destination selection score and enhanced edge content quality (+12\% edge-classification accuracy, +27.9\% BERTScore) over the strongest baseline;
(4) enhanced macro-level fidelity in IDGG tasks, achieving 35.98\% higher structural similarity and 28.71\% better replication of emergent social phenomena (e.g., power laws, echo chambers) when evaluated against ground-truth social graphs.
% (4) demonstrated capability for counterfactual simulation under incentive interventions, enabling exploratory policy analysis.
% In TDGG tasks, interactions are predicted for specific nodes; in IDGG tasks, the model generates the general social graph.

% Guided by these principles, we propose Graphia, a unified reinforcement learning framework for LLM trained on social graph data. 
% Our main contributions are:
% (1) We propose Graphia, a reinforcement learning framework for LLM-based social graph generation that integrates structural feedback via GNN-derived rewards.
% (2) In TDGG experiments, Graphia improves fine-grained social reasoning of LLMs. For destination selection, it improves destination selection by 6.1\% on average over 32B/70B baselines; for edge generation, it improves category accuracy by 12\%, and on text similarity, increases ROUGE-L by 2.5\% and BERTScore by 27.9\%.
% (3) In IDGG experiments, Graphia predict structurally realistic social graphs. Compared to deep-learning and LLM-based graph generators, it improves the structural similarity score by 35.98\% and successfully replicates three social phenomena, with the replication score improved by 28.71\%.
% (4) Graphia supports interventionable simulation, reproducing realistic behavioral shifts (e.g., reposts$\to$ comments) under incentives, enabling macro-level policy analysis.

%  which provide standardized benchmarks for evaluating temporal graph prediction

\section{Related Works}
\subsection{Social Graph Simulation}
Existing social graph simulation methods fall into two categories. Structure-driven models~\cite{TIGGER, DGGen} which capture temporal and topological network dynamics but cannot generate text-rich interactions. LLM-based simulators~\cite{SoMoSiMuBench} generate textual interactions but rely on task-specific pipelines and lack training signals from social graphs. For example, SOTOPIA-RL~\cite{SOTOPIA-RL} trains LLMs using LLM-as-judge rewards on interaction text alone, failing to learn structural properties like degree distributions or homophily. While some works incorporate homophily~\cite{Y_social} or influence propagation~\cite{FUSE} into evaluation, their simulation pipelines are training-free and lack graph-guided learning. This misalignment between training objectives and empirical network structure limits the realism of simulated social dynamics.

\subsection{Social Simulation Evaluation}
LLM-based social simulations are typically evaluated at two levels.
At the micro level, interaction quality is often assessed using the \textit{LLM-as-a-judge} paradigm, which leverages large language models to score dialogue coherence, goal fulfillment, or social appropriateness~\cite{zhousotopia, wangcoser}. While scalable, this approach is sensitive to prompts and exhibits inconsistencies~\cite{llm-judge-problem,zhou2025personaeval}.
At the macro level, current studies rely on qualitative validation of emergent phenomena in specific simulation scenarios, such as attitude shifts~\cite{yao2025social} or information diffusion~\cite{S3} on Twitter. These assessments are tied to specific simulation scenarios or datasets, making them unsuitable for systematic or cross-dataset comparison. Conseqently, a unified quantitative framework that jointly evaluates micro-level interactions and macro-level network realism remains lacking.

\section{Proposed Framework}
We focus on modeling human-like behaviors in dynamic text-attributed social graphs. In this section, we define the social graph data structure, describe the post-training of LLMs for aligning with social behaviors, and introduce our LLM-based framework for social graph simulation.
% We focus on modeling human-like behaviors in large-scale, dynamic social graphs. In this section, we define the social graph data structure, describe the post-training of LLMs for aligning with social behaviors, and introduce our LLM-based framework for social graph generation.

\subsection{Problem Formulation}
\label{sec:problem_formulation}
We consider a directed, dynamic social graph represented as a sequence of time-stamped subgraphs $\{G_t\}_{t=1}^T$, where $G_t = (\mathcal{V}_t, \mathcal{E}_t, \mathbf{P}_t, \mathbf{X}_t)$. 
Here, $\mathcal{V}_t$ is the node set at time $t$, each representing a person; $\mathcal{E}_t \subseteq \mathcal{V}_t \times \mathcal{V}_t$ is directed edge set, denoting interactions from one person to another. 
For node attribute, $\mathbf{P}_t = \{p_v \mid v \in \mathcal{V}_t\}$ contains the node profiles, where $p_v$ is a textual description of node $v$ (e.g., interests, role). 
For edge attribute, $\mathbf{X}_t = \{(m_{u\to v}, y_{u\to v}) \mid (u, v) \in \mathcal{E}_t\}$ is the edge set, where $m_{u\to v}$ is the textual message content and $y_{u\to v} \in \{1, \dots, Y\}$ is its interaction category (e.g., a post or comment).
Given a historical window of length $\tau$, we define the observed sequence as $\mathcal{G}_{\text{hist}} = \mathcal{G}_{1:T-\tau} = \{G_1, \dots, G_{T-\tau}\}$, and the goal is to generate the future sequence $\hat{\mathcal{G}}_{\text{fut}} = \hat{\mathcal{G}}_{T-\tau+1:T} = \{\hat{G}_{T-\tau+1}, \dots, \hat{G}_T\}$.
To capture dynamic behavioral context, we define the node memory $\mathcal{M}_{t}(u)$ for node $u$, which records its past interactions within the historical window:
\[
\mathcal{M}_{t}(u) = \left\{ \big(p_v, m_{u \to v}, y_{u \to v}\big) \,\middle|\, (u, v) \in \mathcal{E}_{<t} \right\}.
\]
This memory includes both the destination nodes' profiles $p_v$ and the previous messages $m_{u\to v}$ with their semantic categories $y_{u\to v}$.

Following GDGB~\cite{peng2025gdgb}, we decompose the social graph simulation task into two settings: 
TDGG and IDGG.
% \textit{Transductive Dynamic Graph Generation (TDGG)} and \textit{Inductive Dynamic Graph Generation (IDGG)}. 
The generation of each interaction in the social graph is modeled as a Markov process:
\begin{equation}
  \label{equa:generation_probab}
  \begin{aligned}
  p(u, v, m, y \mid \mathcal{G}_{\text{hist}}) 
  &= p(u \mid \mathcal{G}_{\text{hist}})
  \cdot p(v \mid u, \mathcal{G}_{\text{hist}})
  \\
  &\hphantom{{} = {}}
  \cdot
  p(m, y \mid u, v, \mathcal{G}_{\text{hist}}).
  \end{aligned}
\end{equation}

In the TDGG task, the active source node set is given. The model estimates $p(v \mid u, \mathcal{G}_{\text{hist}})$ for destination selection and $p(m, y \mid u, v, \mathcal{G}_{\text{hist}})$ for edge generation. This task focuses on micro-level evaluation of interaction patterns between $u$ and $v$.

In the IDGG task, source nodes are not provided; the model must learn $p(u \mid \mathcal{G}_{\text{hist}})$ endogenously. This requires modeling the full generative process of future graph evolution. This task focuses on macro-level evaluation by assessing how well the generated future graph $\hat{\mathcal{G}}_\text{fut}$ reproduces realistic social network structures and dynamic patterns.

\subsection{Graphia Learning Framework}
Building upon Equation~\eqref{equa:generation_probab}, we develop a learning framework to train LLMs for simulating human-like node behaviors in dynamic graphs. Based on the trained LLMs, we design a unified graph generation pipeline for TDGG and IDGG tasks. The overall framework is illustrated in Figure~\ref{fig:framework}.

\noindent \textbf{Activity Prediction.}
To capture which nodes will become active, i.e., $p(u \mid \mathcal{G}_{\text{hist}})$, we introduce the Activity-Predictor, which is implemented with the Informer architecture~\cite{zhou2021informer}. 
For each source node $u\in \mathcal{V}_T$, it takes the historical out-degree sequence $\{d_t(u)\}_{t=1}^{T-\tau}$ as input and predicts the out-degrees over the future horizon: $\{\hat{d}_{T-\tau+1}(u), \dots, \hat{d}_T(u)\}$.
This module is trained to minimize the mean squared error between predicted and actual out-degree:
\begin{equation*}
    \mathcal{L}_{\text{deg}} = \frac{1}{\tau N} \sum_{u \in \mathcal{V}_T} \sum_{t=T-\tau+1}^{T} \left( d_t(u) - \hat{d}_t(u) \right)^2,
\end{equation*}
where $d_t(u)$ denotes the true out-degree of node $u$ at time $t$, and $N=|\mathcal{V}_T|$. The predicted out-degrees serve as structural priors for identifying future active source nodes in the IDGG task.

\noindent \textbf{Interaction Policy Learning.}
For modeling $p(v \mid u, \mathcal{G}_{\text{hist}})$ and $p(m, y \mid u, v, \mathcal{G}_{\text{hist}})$, we treat the LLM as a policy model trained via reinforcement learning. We train two specialized LLMs, Graphia-Q for destination selection and Graphia-E for edge generation, each optimized for its respective task.
% We decompose the policy into two stages: destination selection and edge generation. We train two LLMs, Graphia-Q and Graphia-Q for these two tasks.
% 首先，这边训练的是LLM, named \call{Graphia-Q}. For source node $u$, let $C_t^{u}$ denote the ground-truth destination node set。{Graphia-Q}做的是 predict $\hat{C}_t^u$ destination node set。
% 其次要说明candidate set的选择流程。为了retrieve dst node set $\hat{C}_t^u$，LLM-query generate query and filter rule. （存在两个sample size K1，K2。K1是一个人为设置的常数（100），K2是outdegree）。由filter rule进行初步过滤，we select candidate nodes based on query and historical destination and filtered node profile bert embedding similarity. Duplicates are removed while preserving ranking order. 然后截断到outdegree 个dst node，$K2 = \text{round}(\hat{d}_t(u))$。 details in Appendix~\ref{app:destination_selection}.details in Appendix~\ref{app:destination_selection}.
% 最后要说明的是。为了提升Graphia-Q的准确度，We design a hybrid reward combining structural plausibility and retrieval accuracy:We design a hybrid reward combining structural plausibility and retrieval accuracy:

(1) Destination Selection. For each source node $u$, we train a generative LLM, Graphia-Q, to predict the ground-truth destination node set $C_t^u$ at time $t$. 
We denote the predicted destination node set as $\hat{C}_t^u$.
To retrieve $\hat{C}_t^u$, Graphia-Q generates a descriptive query to constrain the search space. 
% We adopt a two-step destination selection workflow. 
We first retrieve a preliminary candidate set of $K_1$ by ranking nodes in $\mathcal{V}_T$ based on semantic similarity (BERT embedding cosine similarity) to the query, restricted to historical neighbors satisfying the filter rule. 
This leads to the first candidate node set: $\hat{C}_t^{u1}$. 
For graph generation, $ \hat{C}_t^{u1} $ is truncated to size $\hat{C}_t^{u} = ( \hat{C}_t^{u1})_{:\mathrm{round}(\hat{d}_t(u))}$, where $ \hat{d}_t(u) $ is either given (TDGG) or predicted (IDGG). Specifically, following GAD~\cite{GAD}, we observe that common neighbors serve as an effective filter function; thus, we re-rank items retrieved via common neighbors. Detailed process is provided in Appendix~\ref{app:destination_selection}.
To train Graphia-Q, we design a hybrid reward function:
\begin{equation*}
    r_{\text{dst}} = r_{\text{format}} + \alpha\sum_{t=T-\tau+1}^{T}\sum_{v \in C_t^{u}} \mathbb{I}(v \in \hat{C}_t^{u1}),
\end{equation*}
where $ r_{\text{format}} $ is 1 if the generated query conforms to the required format and 0 otherwise, and the second term counts how many true destinations appear in the top-$K_1$ candidates. The hyperparameter $\alpha$ controls the relative strength of the reward for correctly retrieving true destinations.
% where $r_{\text{format}} = 1 $ if both the generated query conform to the required form, and $ 0 $ otherwise. The second term measures retrieval effectiveness by counting the number of true destinations included in the top-$K_1$ candidates.

% We set $ \alpha = 5 $ to strengthen the reward for correct retrievals.

(2) Edge Generation. We train Graphia-E, a generative LLM that generates both the message $\hat{m}_{u\to v}$ and interaction category $\hat{y}_{u\to v}$ for each node pair $(u, v)$. 
To ensure valid output formatting, we include a format reward $r_{\text{format}}$, computed via rule-based parsing, with $r_{\text{format}} = 1$ if valid, else $0$. 
To train Graphia-E, we design separate reward functions for two subtasks: category prediction and message generation. 
For category prediction, we adopt a curriculum-style reward function that measures prediction accuracy, with the emphasis shifting progressively over training epochs:
\begin{equation*}
    r_{\text{cat}} = r_{\text{GNN}} + \beta \cdot \mathbb{I}(\hat{y}_{u\to v} = y^*) + r_{\text{format}},
\end{equation*}
where $\mathbb{I}(\cdot)$ is the indicator function, $r_{\text{GNN}} = [\mathbf{z}_{u,v}]_{y^*} $ is the logit score for the ground-truth category $y^*=y_{u\to v}$ from a pre-trained DGNN edge classifier (We adopt GraphMixer~\cite{graphmixer}), serving as a structural prior to guide the model. We set $\beta = \min(\max(0.01s, \beta_{\text{min}}),\beta_{\text{max}})$, which increases with training step $s$, gradually shifting the reward emphasis from the DGNN’s soft guidance to exact category matching.

For message generation, we adopt the \textit{LLM-as-a-judge} paradigm~\cite{zhousotopia,peng2025gdgb} to assess social and semantic quality. A Qwen3-8B LLM rewarder~\cite{yang2024qwen2technicalreport} scores each generated message on six dimensions: Goal Fulfillment (GF) from SOTOPIA~\cite{zhousotopia}; Contextual Fidelity (CF), Personality Depth (PD), Dynamic Adaptability (DA), Immersive Quality (IQ), and Content Richness (CR) from GDGB~\cite{peng2025gdgb}.
Each dimension is rated on a [0, 5] scale; missing scores are treated as 0.
The final message reward function is normalized and averaged:
\begin{equation*}
    r_{\text{text}} = \frac{1}{6} \sum_{d \in \mathcal{D}} \frac{\text{Score}_d}{5} + r_{\text{format}},
    \label{eq:reward_text}
\end{equation*}
where $ \mathcal{D} = \{\text{GF, CF, PD, DA, IQ, CR}\} $.
For each domain, we define a task-specific reward with shared format regularization. 
Training proceeds via domain-interleaved sampling, with ratio typically 1:1 (category: message). 

During training of Graphia-Q and Graphia-E, we first fine-tune the backbone LLM using SFT, then optimize both tasks with GRPO~\cite{grpo} based on the designed reward function.

\subsection{Graph Generation Pipeline}

\begin{algorithm}[t]
\caption{Graph Generation Pipeline}
\label{alg:generation_pipeline}
\begin{algorithmic}[1]
\Require Historical graph sequence $\mathcal{G}_{\text{hist}} = \{G_1, \dots, G_{T-\tau}\}$, future horizon $\tau$
\Ensure Generated future graph sequence $\hat{\mathcal{G}}_{\text{fut}} = \{\hat{G}_{T-\tau+1}, \dots, \hat{G}_T\}$

\State \textbf{Stage 1: Activity Prediction}
\If{Task is TDGG, $t =\{T-\tau+1,\dots,T\}$}
    \State Given source node set $\mathcal{U}_{t}$,
\Else\ {Task is IDGG, $t =\{T-\tau+1,\dots,T\}$}
    \State Predict out-degrees $\{\hat{d}_t(u), u \in \mathcal{V}_{t}\}$,
    \State Source node set $\mathcal{U}_{t} = \{ u \mid \exists t, \hat{d}_t(u) > 0\}$,
\EndIf

\State \textbf{Stage 2: Interaction Generation}

\For{each $t$ from $T-\tau+1$ to $T$}
    \For{each $u \in \mathcal{U}_{t}$}
        % \State \textbf{Destination Selection:}
        \State \Call{Graphia-Q}{$p_u$, $\mathcal{M}_{t}(u)$} 
        \State \quad = Query, Filter
        \State Retrieve destination nodes $\hat{C}_t^{u}$
        % \State \quad Split $\hat{C}_t$ into time-specific sets $\hat{C}_{T-\tau+1}^{u}, \dots, \hat{C}_T^u$, $|\hat{D}_t^{u}| = \text{round}(\hat{d}_t(u))$

        % \State \textbf{Edge Generation:}
        \For{each $v \in \hat{C}_t^{u}$}
            \State \Call{Graphia-E}{$p_u, p_v$, $\mathcal{M}_{t}(u,v)$} 
            \State \quad = $(\hat{m}_{u\to v}, \hat{y}_{u\to v})$
            \State Add $(u, v, \hat{m}_{u\to v}, \hat{y}_{u\to v})$ to $\hat{\mathcal{E}}_t$
            \State Add $(\hat{m}_{u\to v}, \hat{y}_{u\to v})$ to $\hat{\mathbf{X}}_t$
        \EndFor
    \EndFor
\EndFor

\State Assemble $\hat{\mathcal{G}}_{\text{fut}} = \{ (\mathcal{V}_t, \hat{\mathcal{E}}_t, \mathbf{P}_t, \hat{\mathbf{X}}_t) \}_{t=T-\tau+1}^T$
\State \Return $\hat{\mathcal{G}}_{\text{fut}}$
\end{algorithmic}
\end{algorithm}

We design distinct generation pipelines for TDGG and IDGG to reflect their different evaluation focuses: TDGG emphasizes local agent behaviors
while IDGG targets social network dynamics. As shown in Alg.~\ref{alg:generation_pipeline}, the process involves two stages: activity prediction and interaction generation.

\noindent\textbf{TDGG Pipeline.} \ In the transductive setting, the set of future source nodes is given. The full generation pipeline is:
(1) For given source node $u$, condition Graphia-Q on the node profile $p_u$ and node memory $\mathcal{M}_{t}(u)$ to generate the descriptive query. 
(2) Retrieve the destination node set $\hat{C}_t^{u}$ for source node $u$ with the descriptive query.
(3) For destination node $v \in \hat{C}_t^{u}$, condition Graphia-E on $p_u$, $p_v$, and node memory $\mathcal{M}_{t}(u,v)$ to generate the interaction message $(m_{u \to v}, y_{u\to v})$.

\noindent\textbf{IDGG Pipeline.} \ In the inductive setting, no future source nodes are provided. The full generation pipeline is:
(1) Use the Activity-Predictor to predict out-degrees $\hat{d}_t(v)$ for all nodes.
(2) Select active source nodes with $\hat{d}_t(v) > 0$.
(3) For active source node $u$, apply the Graphia-Q for destination selection and Graphia-E for edge generation.
(4) Assemble the full $\hat{\mathcal{G}}_{\text{fut}} = \{\hat{G}_{T-\tau+1}, \dots, \hat{G}_T\}$.

% For social network generation; TDGG on micro-scale behavioral modeling of single node, relying on provided structural priors, while IDGG focus on macro-scale simulation of emergent social network structure evolution without external guidance. Overall, Graphia thus unifies micro-scale behavioral generation and macro-scale structure prediction.

\section{Experiment}

\subsection{Experimental Setup}
We evaluate both TDGG and IDGG tasks for social graph simulation. In our experiments, we adopt Qwen3-8B~\footnote{https://huggingface.co/Qwen/Qwen3-8B} as the backbone for Graphia.
\label{sec:experiment_setup}

\noindent\textbf{Micro-Level Alignment Metrics.} \ We propose the TDGG score ($S_{\text{TDGG}}$) to evaluate LLM-based agent's local social behavior. 
For destination selection, we measure whether Graphia-Q can identify interaction partners for a source node $u$.
Given ground-truth destination set $C_t^u$ at time $t$, we compute recall at 100 as $\mathrm{R}@100 = |C_t^{u} \cap \hat{C}_t^{u}| / |C_t^u|$, where $\hat{C}_t^{u}$ denotes the top-100 predicted destinations. 
% We categorize each sample by $|C_t^u|$; if $|C_t^u| > \mathrm{70}\text{-th percentile}$, it is \textit{Easy}; otherwise \textit{Hard}. 
We categorize samples into \textit{Easy} and \textit{Hard} based on the size of the ground-truth destination set $|C_t^u|$; if $|C_t^u|$ exceeds the 70-th percentile across all samples, it is labeled \textit{Easy}, otherwise \textit{Hard}.
We report $\mathrm{R}@100$ on Easy, Hard, and All samples. 
Metrics are normalized and aggregated into a summed average $S_{\text{selection}} $.
For edge generation, we assess whether Graphia-E generates valid interaction messages by measuring category prediction accuracy ($\mathrm{ACC}$) of $y_{u \to v}$, and evaluating ROUGE-L and BERTScore-F1 of the generated $\hat{m}_{u \to v}$ against reference message content $m_{u \to v}$.
Metrics are normalized and aggregated into a summed average $S_{\text{edge}}$. The final TDGG score is $S_{\text{TDGG}} = 0.5 \cdot S_{\text{selection}} + 0.5 \cdot S_{\text{edge}}$. 

\begin{table*}[htbp]
\centering
\small
\caption{Evaluation results for destination selection and edge generation in TDGG tasks. The best and second-best results are highlighted in \textbf{bold} and \underline{underline}, respectively.}
\label{tab:tdgg_results}
\begin{adjustbox}{width=\textwidth, totalheight=\textheight, keepaspectratio}
\begin{tabular}{l|rrrrr|rrrrr|rr}
\toprule
Model & \multicolumn{5}{c|}{Destination Selection} & \multicolumn{5}{c|}{Edge Generation} & \multicolumn{2}{c}{TDGG} \\
& $\mathrm{R}@100\text{-Easy}$ $\uparrow$ & $\mathrm{R}@100\text{-Hard}$ $\uparrow$ &$\mathrm{R}@100\text{-All}$ $\uparrow$ & $S_\text{sel}$ $\uparrow$ & Rank 
$\downarrow$ & $\mathrm{ACC}$ $\uparrow$ & ROUGE-L $\uparrow$ & BERTScore $\uparrow$ & $S_\text{edge}$ $\uparrow$ & Rank 
$\downarrow$ & $S_\text{TDGG}$ $\uparrow$ & Rank $\downarrow$ \\
\midrule
\addlinespace
\multicolumn{13}{c}{\textbf{Propagate-En}} \\
\midrule
 Qwen3-8B & 0.4451 & 0.3275 & 0.3634 & 0.2526 & 5.17 & 0.0136 & 0.6810 & 0.6157 & 0.2687 & 4.67 & 0.2606 & 6 \\
 Qwen3-8B-SFT  & 0.4601 & 0.3275 & 0.3677 & 0.5847 & 3.67 & \underline{0.0153} & \underline{0.7332} & \underline{0.7622} & \underline{0.6922} & \underline{2.00} & \underline{0.6385} & \underline{2} \\
 Qwen3-32B & 0.4444 & \underline{0.3415} & 0.3718 & 0.6108 & 3.33 & 0.0088 & 0.6626 & 0.5117 & 0.0000 & 7.00 & 0.3054 & 4 \\
 DeepSeek-Q-32B & \underline{0.4617} & 0.3125 & 0.3582 & 0.1749 & 5.50 & 0.0101 & 0.6668 & 0.5192 & 0.0439 & 6.00 & 0.1094 & 7 \\
 LLama3.1-70B & 0.4418 & \textbf{0.3437} & \underline{0.3735} & \underline{0.6261} & \underline{3.00} & 0.0136 & 0.6978 & 0.6794 & 0.4184 & 3.33 & 0.5223 & 3 \\
 Graphia-seq & 0.4439 & 0.3297 & 0.3641 & 0.2741 & 5.33 & 0.0145 & 0.6824 & 0.6177 & 0.2881 & 3.67 & 0.2811 & 5 \\
 Graphia & \textbf{0.4763} & 0.3319 & \textbf{0.3761} & \textbf{0.8739} & \textbf{1.67} & \textbf{0.0346} & \textbf{0.7421} & \textbf{0.7799} & \textbf{1.0000} & \textbf{1.00} & \textbf{0.9370} & \textbf{1} \\
\midrule
\addlinespace
\multicolumn{13}{c}{\textbf{Weibo Tech}} \\
\midrule
 Qwen3-8B & 0.2602 & 0.2301 & 0.2455 & 0.5612 & 4.50 & 0.6326 & \underline{0.6014} & 0.1757 & 0.5367 & 3.00 & \underline{0.5490} & \underline{2} \\
 Qwen3-8B-SFT  & 0.2422 & 0.2250 & 0.2334 & 0.0455 & 6.33 & \underline{0.7325} & 0.5985 & \underline{0.2128} & \underline{0.6920} & \underline{2.67} & 0.3687 & 6 \\
 Qwen3-32B & 0.2641 & \underline{0.2346} & 0.2498 & \underline{0.8138} & \underline{2.50} & 0.5765 & 0.5762 & 0.0873 & 0.0314 & 6.67 & 0.4226 & 5 \\
 DeepSeek-Q-32B & \textbf{0.2767} & 0.2235 & \underline{0.2518} & 0.6579 & 2.83 & 0.5846 & 0.5790 & 0.0656 & 0.0449 & 6.33 & 0.3514 & 7 \\
 LLama3.1-70B & 0.2607 & 0.2283 & 0.2448 & 0.5256 & 4.83 & 0.6453 & 0.5929 & 0.1459 & 0.4095 & 4.33 & 0.4675 & 4 \\
 Graphia-seq & 0.2629 & 0.2232 & 0.2438 & 0.4146 & 5.33 & 0.6297 & 0.6007 & 0.1747 & 0.5230 & 4.00 & 0.4688 & 3 \\
 Graphia & \underline{0.2700} & \textbf{0.2364} & \textbf{0.2538} & \textbf{0.9292} & \textbf{1.50} & \textbf{0.8221} & \textbf{0.6040} & \textbf{0.2963} & \textbf{1.0000} & \textbf{1.00} & \textbf{0.9646} & \textbf{1} \\
\midrule
\addlinespace
\multicolumn{13}{c}{\textbf{Weibo Daily}} \\
\midrule
 Qwen3-8B & 0.3326 & 0.3030 & 0.3135 & 0.6719 & 4.83 & 0.5456 & 0.5661 & \underline{0.1243} & 0.4201 & 2.67 & 0.5460 & 3 \\
 Qwen3-8B-SFT  & 0.3096 & 0.3060 & 0.3072 & 0.3033 & 5.50 & \underline{0.6338} & \underline{0.5755} & 0.0735 & \underline{0.4625} & 2.67 & 0.3829 & 6 \\
 Qwen3-32B & 0.3344 & \textbf{0.3159} & \textbf{0.3226} & \textbf{0.9362} & \textbf{1.50} & 0.3921 & 0.5332 & 0.0142 & 0.0438 & 6.67 & 0.4900 & 4 \\
 DeepSeek-Q-32B & \textbf{0.3401} & 0.2769 & 0.2997 & 0.3675 & 4.83 & 0.4127 & 0.5332 & -0.0238 & 0.0140 & 6.33 & 0.1908 & 7 \\
 LLama3.1-70B & 0.3259 & \underline{0.3127} & \underline{0.3173} & \underline{0.7470} & 3.33 & 0.5332 & 0.5450 & 0.0490 & 0.2315 & 5.00 & 0.4892 & 5 \\
 Graphia-seq & 0.3325 & 0.3042 & 0.3142 & 0.6844 & 4.33 & 0.5422 & 0.5657 & 0.1224 & 0.4138 & 3.67 & \underline{0.5491} & \underline{2} \\
 Graphia & \underline{0.3379} & 0.3042 & 0.3162 & 0.7411 & \underline{3.17} & \textbf{0.8836} & \textbf{0.6088} & \textbf{0.2652} & \textbf{1.0000} & \textbf{1.00} & \textbf{0.8706} & \textbf{1} \\
\midrule
\addlinespace
 \multicolumn{13}{c}{\textbf{Average Performance}} \\
 \midrule
  \multicolumn{1}{l|}{Best Baseline} & - & - & - & 0.7869 & 2.44 & - & - & - & 0.6156 & 2.44 & - & - \\
 \multicolumn{1}{l|}{Graphia} & - & - & - & \textbf{0.8481} & \textbf{2.11} & - & - & - & \textbf{1.0000} & \textbf{1.00} & - & - \\
\bottomrule
\end{tabular}
\end{adjustbox}
\end{table*}

\noindent\textbf{Macro-Level Alignment Metrics.}
% We design the IDGG social fidelity score ($S_{\text{IDGG}}$) to evaluate how well generated graphs replicate global topology and emergent phenomena in $\mathcal{G}_{\text{fut}}$.
We propose the IDGG score ($S_{\text{IDGG}}$) to evaluate Graphia for predicting social graph structure and emergent social phenomena.
For structure replication, we use Maximum Mean Discrepancy (MMD) with an RBF kernel to measure distributional distances in degree, clustering, and spectral properties~\cite{peng2025gdgb}. We also compute edge overlap, $
\mathrm{EO} = |\hat{\mathcal{E}}_{\text{fut}} \cap \mathcal{E}_{\text{fut}}|/{|\mathcal{E}_{\text{fut}}|}.$ 
Metrics are normalized and aggregated into a summed average $S_{\text{structure}}$.
For phenomenon replication, we evaluate:
(i) Influencer identification. Following SaGraph~\cite{sagraph_bench}, we report $\mathrm{P}@100\text{-KOL}$: the precision of KOLs in the top-100 degree nodes of $\hat{\mathcal{G}}_\text{fut}$.
(ii) Echo chamber alignment. We measure the echo-chamber count difference $\Delta C$~\cite{del2016spreading} between $\mathcal{G}_\text{fut}$ and $\hat{\mathcal{G}}_\text{fut}$.
(iii) Power-law fitness. We measure the power-law exponent gap $\Delta \alpha = |\alpha_{\text{ref}} - \alpha_{\text{gen}}|$, where $\alpha_{\text{ref}}$ and $\alpha_{\text{gen}}$ are fitted exponents for $\mathcal{G}_\text{fut}$ and $\hat{\mathcal{G}}_\text{fut}$~\cite{peng2025gdgb}.
Metrics are normalized and aggregated into a summed average $S_{\text{phenomenon}}$.
The final IDGG score is $S_{\text{IDGG}} = 0.5 \cdot S_{\text{structure}} + 0.5 \cdot S_{\text{phenomenon}}$.

% Following GDGB~\cite{peng2025gdgb}, we assess how well the generated graph preserves the heavy-tailed nature of real-world networks.
%   We fit a power-law distribution $p(k) \sim k^{-\alpha}$ to the degree sequence of the reference graph using maximum likelihood estimation, with $x_{\min} = 2$ for all datasets.
%   The goodness-of-fit is evaluated via the Kolmogorov-Smirnov (KS) distance between the empirical and fitted distributions.
%   Additionally, we compute the deviation in the estimated power-law exponent:
%   \begin{align*}
%       \Delta \alpha = |\alpha_{\text{ref}} - \alpha_{\text{gen}}|,
%   \end{align*}

\noindent\textbf{Baseline Models.}
% This enables ablation on structural modeling.
To assess the impact of graph-structured data on training LLMs, we construct sequential data that retains only first-order neighbor edges while discarding higher-order topology.
For the TDGG task, which focuses on micro-level behavioral alignment, we compare Graphia against a range of LLMs of varying scale: Qwen3-8B, Qwen3-32B, DeepSeek-R1-Distill-Qwen-32B, and Llama-3.1-70B-Instruct, as well as a fine-tuned Qwen3-8B baseline (Qwen3-SFT). 
We further evaluate Graphia-seq, a variant of Graphia trained on sequential interaction data based on Qwen3-8B, using destination selection rewards from LIKR~\cite{LIKR} and edge generation rewards from Sotopia~\cite{zhousotopia}.
For the IDGG task, we compare Graphia with deep-learning and LLM-based social graph generators. First, we adopt DGGen~\cite{DGGen} and TIGGER~\cite{TIGGER}, the only deep-learning models designed for inductive dynamic graph generation; we further note that DGGen uses TGN~\cite{TGN} as its backbone to learn temporal graph embeddings. Following the DGGen framework, we replace the backbone with more recent temporal graph learning models~\cite{DTGB}, yielding DGGen variants with GraphMixer~\cite{graphmixer}, CAWN~\cite{CAWN}, and Dygformer~\cite{Dygformer} backbones. Second, we construct two hybrid LLM-based simulators by pairing Qwen3-SFT and Graphia-seq with the SA-Graph activity-prediction module~\cite{TIS}, which fits Gaussian distributions to activity patterns and samples active nodes. Finally, we include GAG-General~\cite{peng2025gdgb} with Llama3-8B and Qwen3-8B backbones, an LLM-based multi-agent framework for graph generation.

\begin{table}[h]
\centering
\caption{Key statistics of the social network datasets.}
\label{tab:dataset_main_info}
\begin{adjustbox}{width=\linewidth, keepaspectratio}
\begin{tabular}{lrrr}
\toprule
\textbf{Dataset} & \textbf{Propagate-En} & \textbf{Weibo Tech} & \textbf{Weibo Daily} \\
\midrule
Node Count                    & 5,634     & 20,768    & 66,501     \\
Edge Count                    & 16,962    & 58,930    & 195,202    \\
Input (Days)           & 18        & 5         & 19         \\
Prediction (Days)     & 4         & 1         & 4          \\
Total (Days)       & 30        & 8         & 31         \\
Category Number        & 534       & 2         & 2          \\
\bottomrule
\end{tabular}
\end{adjustbox}
\end{table} 

\noindent\textbf{Datasets.} \ We evaluate our framework on three social network datasets: Propagate-En, collected from the Taobao e-commerce platform, and two public datasets: Weibo Tech and Weibo Daily from GDGB~\cite{peng2025gdgb}. 
Table~\ref{tab:dataset_main_info} summarizes the dataset statistics, which are aggregated by day based on edge timestamps. 
Let $ T_2 $ be the total time length and $ \tau = \lfloor 0.15 \times T_2 \rfloor $. The input length is set to $ T-\tau = T_2 - 3\tau $, and the prediction length is $ \tau $. Training, validation, and test sets are chronologically partitioned into intervals of $[0, T]$, $[\tau, T + \tau]$, and $[2\tau, T + 2\tau]$. Detailed dataset statistics and preprocessing are provided in Appendix~\ref{app:dataset_details}.

% LLM as judge
% Inspired by \textbf{SOTOPIA-EVAL}~\cite{zhousotopia}, we adopt the \textit{LLM-as-a-judge} paradigm to assess interaction quality. Using an LLM evaluator (Qwen2-70B~\cite{yang2024qwen2technicalreport}), we score each generated message on six dimensions: Goal Fulfillment (GF) from SOTOPIA~\cite{zhousotopia}, and Contextual Fidelity (CF), Personality Depth (PD), Dynamic Adaptability (DA), Immersive Quality (IQ), and Content Richness (CR) from GDGB~\cite{peng2025gdgb}. Each dimension is scored on a [0,5] scale, with missing scores treated as 0. We report the average score following Equa.~\ref{equa:reward_text}. Higher scores indicate more socially coherent and contextually appropriate interactions.

\begin{table*}[htbp]
  \caption{Evaluation results for social graph structure and social phenomenon replication in IDGG tasks.}
  \label{tab:idgg_results}
  \centering
  \begin{adjustbox}{width=\textwidth, totalheight=\textheight, keepaspectratio}
  \begin{tabular}{l|rrrrrr|rrrrr|rr}
    \toprule
    Model & \multicolumn{6}{c|}{Macro Structure} & \multicolumn{5}{c|}{Macro Phenomenon} & \multicolumn{2}{c}{IDGG} \\
     & $\mathrm{MMD.D}^2$ $\downarrow$ & $\mathrm{MMD.C}^2$ $\downarrow$ & $\mathrm{MMD.S}^2$ $\downarrow$ & EO $\uparrow$ & $S_\text{structure}$ $\uparrow$ & Rank $\downarrow$ & $\mathrm{P}@100\text{-KOL}$ $\uparrow$ & $\Delta C$ $\downarrow$ & $\Delta \alpha$ $\downarrow$ & $S_\text{phenomenon}$ $\uparrow$ & Rank $\downarrow$ & $S_\text{IDGG}$ $\uparrow$ & Rank $\downarrow$ \\
    \midrule
    \addlinespace
    \multicolumn{14}{c}{\textbf{Propagate-En}} \\
    \midrule

    Qwen3-8B-sft & 0.3509 & 0.4128 & 0.3739 & \underline{0.0608} & 0.4054 & 5.25 & 0.27 & 33.0000 & 1.0884 & 0.3054 & 4.67 & 0.3554 & 5.00 \\

    DGGen & \underline{0.1486} & \textbf{0.3336} & \textbf{0.1327} & 0.0000 & \underline{0.6602} & \underline{2.50} & 0.02 & \underline{13.0000} & 0.1614 & \underline{0.5283} & \underline{3.00} & \underline{0.5942} & \underline{2.00} \\
    DGGen(GraphMixer) & 0.1601 & 0.4315 & 0.3253 & 0.0018 & 0.4531 & 5.00 & 0.00 & 30.0000 & 1.1210 & 0.0862 & 6.00 & 0.2697 & 7.00 \\
    DGGen(DyGFormer) & 0.1801 & 1.4142 & 0.4006 & 0.0000 & 0.1352 & 6.75 & 0.02 & 18.0000 & 1.3355 & 0.1793 & 5.00 & 0.1573 & 8.00 \\
    DGGen(CAWN) & 0.1495 & 0.3753 & 0.3158 & 0.0073 & 0.4967 & 3.75 & 0.01 & 31.0000 & 1.2745 & 0.0458 & 6.33 & 0.2713 & 6.00 \\
    Tigger & 0.2067 & 1.3563 & 0.2613 & 0.0000 & 0.2575 & 5.50 & 0.01 & 21.0000 & \underline{0.0216} & 0.4685 & 4.00 & 0.3630 & 3.00 \\
    Graphia-seq & 0.3406 & \underline{0.3522} & 0.3797 & \underline{0.0608} & 0.4222 & 4.50 & \underline{0.28} & 33.0000 & 1.1728 & 0.2932 & 5.00 & 0.3577 & 4.00 \\
    Graphia & \textbf{0.0351} & 0.3557 & \underline{0.1981} & \textbf{0.1022} & \textbf{0.9339} & \textbf{1.75} & \textbf{0.37} & \textbf{2.0000} & \textbf{0.0100} & \textbf{1.0000} & \textbf{1.00} & \textbf{0.9669} & \textbf{1.00} \\
    \midrule

    \addlinespace
    \multicolumn{14}{c}{\textbf{Weibo Tech}} \\
    \midrule
    Qwen3-8B-sft & 0.2623 & 1.2628 & 0.4772 & 0.0143 & 0.2603 & 4.25 & \underline{0.30} & 16.0000 & 1.0828 & \underline{0.8186} & \underline{3.67} & \underline{0.5395} & \underline{2.00} \\
    DGGen & 0.1870 & 1.2979 & 0.5472 & 0.0001 & 0.2363 & 5.25 & 0.01 & 18.0000 & 0.5434 & 0.5909 & 4.67 & 0.4136 & 5.00 \\
    DGGen(GraphMixer) & \underline{0.1292} & 1.4142 & 0.5636 & 0.0000 & 0.2236 & 5.75 & 0.01 & 16.0000 & 1.1252 & 0.5102 & 4.67 & 0.3669 & 6.00 \\
    GAG-General(Qwen3-8B) & 0.4422 & 1.4141 & 0.3503 & 0.0000 & 0.1157 & 6.00 & 0.00 & 112.0000 & 0.4446 & 0.2851 & 5.67 & 0.2004 & 8.00 \\
    GAG-general(Llama3-8B) & \textbf{0.0922} & \underline{1.1131} & 0.1904 & 0.0000 & \underline{0.5687} & \underline{2.75} & 0.00 & 66.0000 & \underline{0.1889} & 0.4753 & 5.00 & 0.5220 & 4.00 \\
    Tigger & 0.3349 & 1.2840 & \underline{0.1727} & 0.0000 & 0.3390 & 4.75 & 0.00 & \underline{14.0000} & 2.3443 & 0.3234 & 5.33 & 0.3312 & 7.00 \\
    Graphia-seq & 0.2692 & 1.2537 & 0.4948 & \underline{0.0144} & 0.2496 & 4.25 & 0.27 & 16.0000 & 1.0113 & 0.7981 & \underline{3.67} & 0.5239 & 3.00 \\
    Graphia & 0.1467 & \textbf{0.7668} & \textbf{0.1027} & \textbf{0.1347} & \textbf{0.9611} & \textbf{1.50} & \textbf{0.32} & \textbf{11.0000} & \textbf{0.1230} & \textbf{1.0000} & \textbf{1.00} & \textbf{0.9805} & \textbf{1.00} \\
    \midrule

    \addlinespace
    \multicolumn{14}{c}{\textbf{Weibo Daily}} \\
    \midrule
    Qwen3-8B-sft & 0.3234 & 0.8353 & 0.4558 & 0.0253 & 0.4444 & 4.25 & \textbf{0.44} & \textbf{0.0000} & 1.0467 & \underline{0.6830} & \textbf{2.67} & \underline{0.5637} & \underline{2.00} \\
    DGGen & 0.2126 & 0.9267 & 0.7379 & 0.0000 & 0.3087 & 5.00 & 0.00 & \underline{1.0000} & 0.3315 & 0.5815 & 3.00 & 0.4451 & 6.00 \\
    GAG-General(Qwen3-8B) & 0.5698 & 1.4142 & 0.3392 & 0.0000 & 0.1416 & 5.50 & 0.00 & 123.0000 & 0.7887 & 0.1010 & 5.33 & 0.1213 & 7.00 \\
    GAG-general(Llama3-8B) & \underline{0.1362} & \underline{0.7363} & 0.2065 & 0.0000 & \underline{0.5869} & \underline{2.75} & 0.00 & 4.0000 & 0.5478 & 0.5025 & 4.33 & 0.5447 & 3.00 \\
    Tigger & 0.2098 & 1.4102 & \underline{0.0648} & 0.0000 & 0.4171 & 3.75 & 0.00 & 8.0000 & \textbf{0.0802} & 0.6450 & 3.67 & 0.5310 & 5.00 \\
    Graphia-seq & 0.3342 & 0.7786 & 0.4735 & \underline{0.0262} & 0.4506 & 4.25 & \underline{0.40} & 3.0000 & 1.0966 & 0.6282 & 4.00 & 0.5394 & 4.00 \\
    Graphia & \textbf{0.0614} & \textbf{0.4983} & \textbf{0.0338} & \textbf{0.0973} & \textbf{1.0000} & \textbf{1.00} & 0.32 & 3.0000 & \underline{0.2074} & \textbf{0.8593} & \textbf{2.67} & \textbf{0.9296} & \textbf{1.00} \\
    \midrule

    \addlinespace
    \multicolumn{14}{c}{\textbf{Average Performance}} \\
    \midrule
    Best Baseline & - & - & - & - & 0.6052 & 2.66  & - & - & - & 0.6766  & 3.11 & - & - \\
    Graphia & - & - & - & -  & \textbf{0.9650} & \textbf{1.42} & - & - & - & \textbf{0.9531} &  \textbf{1.56} & - & - \\
    \bottomrule
  \end{tabular}
  \end{adjustbox}
\end{table*}

\subsection{TDGG: Micro-Level Alignment}
\noindent\textbf{Destination Selection.} \ As shown in Table~\ref{tab:tdgg_results} on the left side, Graphia achieves competitive performance in destination selection. 
First, ablation studies validate the effectiveness of our training framework. Both Qwen3-8B and Qwen3-8B-SFT underperform compared to Graphia. Graphia achieves an average rank of 2.11, outperforming Qwen3-8B-SFT by a margin of 3.
Second, despite being built on an 8B-parameter backbone, Graphia achieves performance comparable to or even exceeding that of larger LLMs. On Propagate-En, it outperforms Qwen3-32B (rank: 3.33), DeepSeek-Q-32B (rank: 5.50), and Llama3.1-70B (rank: 3.00), achieving the highest average rank of 1.67.
While on Weibo Daily, Graphia ranks behind Qwen3-32B; In the aggregated selection score, Graphia achieves an average of $S_\text{sel} = 0.848$ across datasets, surpassing the best baseline Qwen3-32B by 6.1\% ($S_\text{sel} = 0.787$). 
% These results indicate that alignment with graph data enable Graphia to match or exceed the performance of larger LLMs in destination selection.
These results demonstrate Graphia's superior micro-level alignment capabilities, matching or exceeding the performance of much larger models in destination selection.
\begin{figure}[htbp]
  \subfloat{\includegraphics[width=\linewidth]{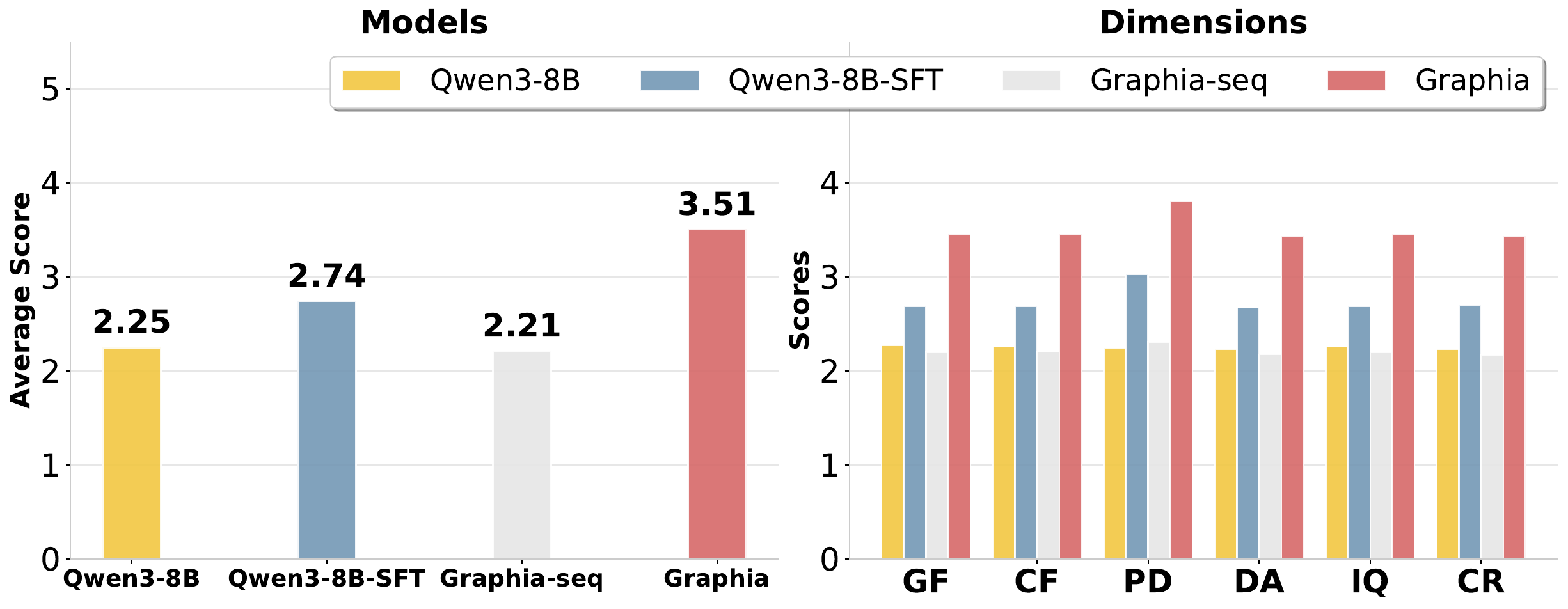
  }}
  % \vspace{1em}
\caption{LLM-as-a-judge for edge generation.}
\label{fig:edge_llm_rating}
\end{figure}

\noindent\textbf{Edge Generation.} \ Inspired by SOTOPIA, we first adopt the LLM-as-a-judge for edge evaluation. Using Qwen2-72B as the evaluator, $m_{u\to v}$ is scored across six dimensions~\cite{yang2024qwen2technicalreport}: \textit{Goal Fulfillment} (GF) from SOTOPIA~\cite{zhousotopia}, and \textit{Contextual Fidelity} (CF), \textit{Personality Depth} (PD), \textit{Dynamic Adaptability} (DA), \textit{Immersive Quality} (IQ), \textit{Content Richness} (CR) from GDGB~\cite{peng2025gdgb}. Ratings range from 1 to 5 (missing values treated as zero).
As shown in Figure~\ref{fig:edge_llm_rating}, we evaluate four models: Qwen3-8B, Qwen3-8B-SFT, Graphia-seq, and Graphia. While Graphia-seq uses the SOTOPIA-RL~\cite{SOTOPIA-RL} reward framework: using LLM-as-a-judge scoring \textit{Goal Fulfillment}, \textit{Relationship Maintenance}, and \textit{Knowledge Seeking} for GRPO training; we find this reward design for sequential data does not improve LLM-judge scores on generated edge messages. In contrast, Graphia achieves the highest average score, outperforming the best baseline by 0.77 points (+28\% relative), with the largest gain in \textit{Personality Depth} by 0.78 points (+25.7\% relative). 
% Moreover, we abalate on different LLMs to judge the txtual content quality, including Qwen3-32B, Qwen2-72B LLama3.1-70B and Llama3.3-70B. 
% Since LLM-as-a-judge is vulnerable to manipulation~\cite{llm-judge-problem}. We The results consistently show that Graphia outperforms baselines across different LLM judges (detailed results in Appendix~\ref{appendix:llm_judge_variants}).
Since LLM-as-a-judge is vulnerable to manipulation~\cite{llm-judge-problem}, we evaluate Graphia with four different LLMs as judge; results consistently show it outperforms baselines (see Appendix~\ref{appendix:llm_judge_variants}, include Qwen3-32B, Qwen2-72B LLama3.1-70B and Llama3.3-70B).

% Therefore, we complement it with task-grounded metrics. 
To ensure a more robust evaluation, we complement LLM-as-a-judge scores with automatic metrics, specifically category accuracy (ACC) for $\hat{y}_{u\to v}$ and content similarity metrics for $\hat{m}_{u\to v}$.  These metrics are then normalized and aggregated into $S_\text{edge}$.
% We evaluate the generated edge $\hat{e}_{u\to v}$ based on category $\hat{y}_{u\to v}$ and content $\hat{m}_{u\to v}$, and then aggregate the metric results into the edge score $S_\text{edge}$.
As shown in Table~\ref{tab:tdgg_results}, Graphia consistently leads on edge generation metrics across datasets. For category accuracy, Graphia improves over the best baseline (Qwen3-8B-SFT) by 1.9\% on Propagate-En, 24.98\% on Weibo Daily, and 9\% on Weibo Tech, resulting in an average improvement of 12\% in category prediction accuracy.
For message content similarity, Graphia improves ROUGE-L by 0.016 points (+2.5\% relative). 
More notably, it improves BERTScore by an average increase of 0.098 points (+27.9\% relative). Graphia maintains a rank of 1 across all evaluation metrics.
In the combined metric $S_{\text{edge}}=1$, Graphia achieves $S_{\text{edge}}=1$ on average, surpassing the best baseline by 38.4\% ($S_{\text{edge}}=0.6156$).
% LLM rating 放一个总的plot图
% 所有数据集的rating图放附录 (雷达图，6个维度)

%% todo
%% 这边放一个average和goal的结果，完整的放附录

% 添加
% key findings:
% 1. Inspired by \textbf{SOTOPIA-EVAL}~\cite{zhousotopia}, we adopt the \textit{LLM-as-a-judge} paradigm to assess interaction quality. Using an LLM evaluator (Qwen2-72B~\cite{yang2024qwen2technicalreport}), we score each generated message on six dimensions: Goal Fulfillment (GF) from SOTOPIA~\cite{zhousotopia}, and Contextual Fidelity (CF), Personality Depth (PD), Dynamic Adaptability (DA), Immersive Quality (IQ), and Content Richness (CR) from GDGB~\cite{peng2025gdgb}. Each dimension is scored on a [0,5] scale, with missing scores treated as 0. 对于Qwen3-8B、Qwen3-8B-SFT、Graphia-seq、Graphia进行了评价。其中Graphia-seq的reward follow了Sotopia的reward设计~\cite{zhousotopia}，均为LLM-as-judge对于Goal Completion、Relationship Maintenance (REL)、Knowledge Seeking (KNO)进行打分。
% 首先发现Sotopia的reward设计~\cite{zhousotopia}在sequential data上数据，无法提升在graph edge message的llm打分。
% 我们发现Graphia在ersonality Depth维度上提升最多,相比Best baseline提升了0.78分（16%）。并且在average相比Best baseline提升了0.77分（15%）。
% 2. 有工作指出，LLM-as-judge is easy to hack。因而我们认为在sicial simulation上，仅仅依赖llm rating是不够的。As shown in Table~\ref{tab:tdgg_results}，在category分类上，Graphia相比最佳baseline，在($y_{u\to v}$)category classification上平均提升...。而在message content$m_{u\to v}$ 的文本相似度指标ROUGE-L和BERTScore上平均提升。。。，
% 3. 综合category和message的指标，我们发现在edge score ($S_\text{edge}$)上，由于Graphia稳定的在各个metric保持第一，因而$S_\text{edge}$ =1，相比best baseline（Qwen3-8B-SFT）提升了。。。。

% \vspace{-2ex}

% \vspace{-2ex}
\subsection{IDGG: Macro-Level Alignment}
\noindent\textbf{Macro Structure Replication.} \ We evaluate all baselines on the three datasets, omitting results for those that encounter out-of-memory (OOM) errors. As shown in Table~\ref{tab:idgg_results}, Graphia consistently attains the lowest $\mathrm{MMD.D}^2, \mathrm{MMD.C}^2$ and $\mathrm{MMD.S}^2$ scores across all datasets. While DGGen performs competitively on Propagate-En, LLM-based generators, including Graphia, consistently outperform deep learning baselines on both Weibo datasets. Notably, most deep learning models yield near-zero edge overlap (EO), reflecting a significant gap in edge distribution from reference graphs. In contrast, LLM-based methods generate non-zero EO, with Graphia achieving the highest EO on all datasets. Averaged across datasets, Graphia attains $S_\text{structure} = 0.97$, surpassing the best baseline by 35.98\% ($S_\text{structure} = 0.61$).

\begin{figure}[htbp]
  \centering
  \subfloat[TDGG Evaluation]   
    {
        \centering
        \includegraphics[width=.5\linewidth]{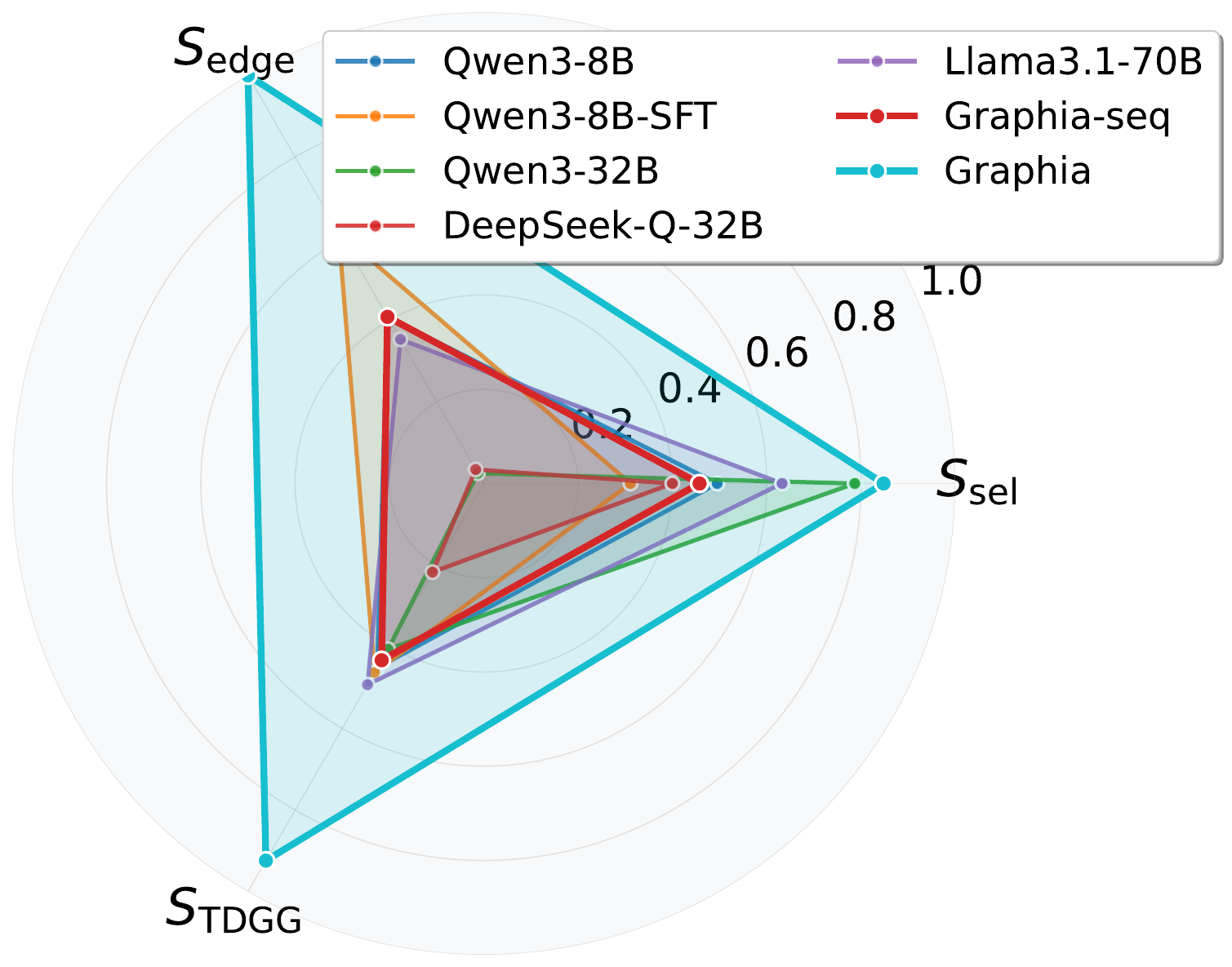}
        \label{fig:powerlaw_reason_d}
    }
  \subfloat[IDGG Evaluation]  
    {
        \centering
        \includegraphics[width=.5\linewidth]{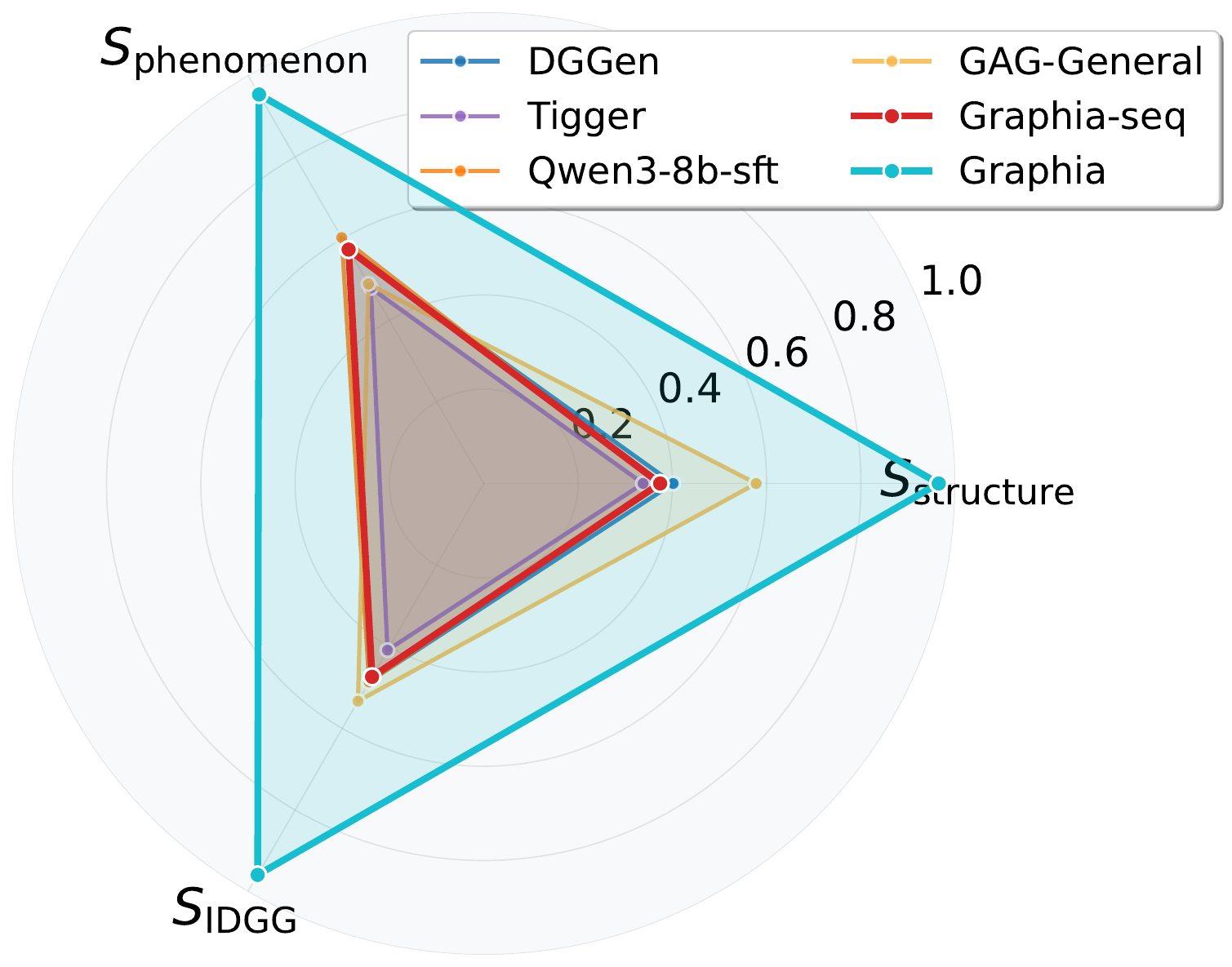}
        \label{fig:powerlaw_reason_mi}
    }
    \caption{The social fidelity score for TDGG and IDGG tasks. 
    % Notably, Graphia exceeds Graphia-seq across all metrics, underscoring the necessity of graph data for enhancing LLM-based social graph simulation.
    % (a) In TDGG tasks, Graphia outperforms baselines in edge generation and matches 32B models in destination selection; (b) In IDGG tasks, Graphia outperforms deep learning-based graph generators, hybrid, and pure LLM-based simulators. We select GAG-General with better-performing Llama3 backbone.
    (a) Graphia outperforms baseline in edge generation, achieves equal performance with 32B in destination selection. (b) Graphia outperforms baselines in graph structure and phenomena replication, achieves best performance compared to both deep-learning based and LLM-based social graph generators.
}
\end{figure}

\noindent\textbf{Macro Phenomenon Replication.} \ To quantitatively assess emergent societal phenomena, we introduce three metrics: $\mathrm{P}@100\text{-KOL}$, echo chamber alignment ($\Delta C$), and deviation in power-law exponent ($\Delta \alpha$). These metrics address the limitations of existing LLM-based simulators, which rely on qualitative assessments of phenomena such as echo chambers~\cite{zheng2024simulating,coling_echo_chamber}, power-law distributions~\cite{GraphMaster,GAG}, and influencer selection~\cite{sagraph_bench}. Graphia achieves the best performance in these metrics on Weibo-Tech and Propagate-En, and ranks 2.67 in Weibo-Daily. Averaged across datasets, Graphia achieves $S_\text{phenomenon}=0.95$, surpassing the best baseline by 28.71\% ($S_\text{phenomenon} = 0.68$), suggesting it's reliable for exploration of phenomena in social graphs.

\section{Conclusion}
% In this paper, we study the alignment of LLM-based social simulations with social network dynamics through reinforcement learning on social graph data.
% By incorporating GNN-derived structural feedback into the reward function, our proposed framework, Graphia, aims to improve LLM-based social graph simulators in terms of both micro-level interaction realism and macro-level social network coherence. 
% Experiments on three real-world social networks show that, at the micro-level for interaction pattern alignment, Graphia outperforms strong baselines in destination selection (+6.1\% in the composite destination selection score) and edge generation (+12\% in category accuracy, +27.9\% in BERTScore). 
% At the macro-level for social graph alignment, the generated graphs exhibit improved structural fidelity, with a 35.98\% higher graph structure similarity; along with 28.71\% better replication of emergent social phenomena in the generated graphs, including power-law degree distribution, echo chambers, and influencer dynamics.
% Additionally, Graphia supports counterfactual simulation: when exposed to platform-level incentives, the generated social graphs produce directionally plausible shifts in interaction patterns. These results suggest that integrating graph structure into LLM training can help bridge local behavior modeling and global network coherence, offering a path toward more realistic and analytically useful social graph simulation.

In this paper, we address two critical limitations in LLM-based social simulation: (i) the absence of a generalizable training framework that leverages graph-structured data to enhance microscopic and macroscopic social simulation realism, and (ii) the lack of unified quantitative metrics to assess the alignment between simulated and real-world social graphs.
To bridge these gaps, we make two key contributions. First, we propose Graphia, a general social graph generator that treats social graph as high-quality supervision signals for LLM post-training. 
Graphia trains specialized LLM-based agents to model human-like interactions by predicting whom to interact and how to interact, followed by carefully designed graph generation pipelines. Second, we establish a unified evaluation paradigm for TDGG and IDGG tasks, with quantitative metrics to assess both micro-level interaction and macro-level realism in social graph simulation. 
Experiments on three real-world datasets validate both contributions: Graphia improves micro-level alignment by 6.1\% in destination selection, 12\% in edge classification accuracy, and 27.9\% in edge content BERTScore; simultaneously, it achieves 35.98\% higher structural similarity and 28.71\% better replication of emergent social phenomena for macro-level alignment. 
This shows that social graphs are effective supervision signals for LLMs, bridging microscopic agent behaviors and macroscopic network dynamics in social simulation.
% These results suggest that integrating graph structure into LLM training can help bridge local behavior modeling and global network coherence, offering a path toward more realistic and analytically useful social graph simulation.

\section*{Limitations}
This paper acknowledges several limitations that future research could address:

\noindent\textbf{Analysis of Learned Policies.}  Our primary focus is improving alignment between LLM-generated and real-world social graph simulation via reward feedback from social graphs. Our study focuses on who, how, and when agents interact within the simulated social network. The question of why agents interact remains outside the scope of this work.Yet, a core tenet of social network analysis is that agent behaviors stem from interpretable causal mechanisms, and that these micro-level mechanisms give rise to distinct macro-level network phenomena. Future work can build upon Graphia’s framework to explicitly investigate the causal drivers of both micro-level agent decisions and emergent macro-level graph structures. 
% Nevertheless, the high-fidelity simulations produced by Graphia provide a robust foundation for such causal inquiries in subsequent research.
% \noindent\textbf{Constrained Counterfactuals.} \ Our primary focus is improving alignment between LLM-generated and real-world social graph simulation via structural feedback from social graphs. Accordingly, we evaluate only two platform-wide incentives, applied as one-shot broadcasts without variation in timing, duration, or intensity. We leave longitudinal policy analysis to future work. Our study focuses on who, how, and when agents interact within the simulated social network. The question of why agents interact remains outside the scope of this work. Nevertheless, the high-fidelity simulations of Graphia establish a robust foundation for investigating these deeper behavioral and cognitive dimensions in subsequent research.
% While the results exhibit plausible short-term shifts in interaction composition, we do not analyze longer-term or cascading effects (e.g., evolving engagement, diffusion dynamics). 

% We leave longitudinal policy analysis to future work. Nonetheless, the high-fidelity simulations enabled by Graphia provide a solid foundation for such studies.

\noindent\textbf{Incorporation of Structural Rewards.} \ While our framework uses GNN-derived rewards to align LLM-generated edges with real-world graph edges, the current implementation relies solely on node-level destination selection and edge-category logits from a pre-trained dynamic GNN. Higher-order topological properties (such as community cohesion, triadic closure) are not explicitly captured in the reward function.  
% We have also experimented with incorporating GNN-based link prediction scores directly into the destination selection reward; however, this approach proved less robust than the overlap-based reward on large-scale graphs.  
Future work could investigate learned filtering mechanisms or graph-structure-aware reward designs that explicitly incorporate complex topological signals, enabling better generalization across diverse graph regimes.
% Additionally, the filter rules used by Graphia-Q are manually specified rather than learned, which may affect adaptability across diverse network domains. 

%% 本篇论文为social graph generation提供了一个统一的量化框架：micro and macro alignment，为llm-based social simulation的真实度对应给出了两个量化指标。为基于深度学习和基于LLM-based social graph simulator提供了可比较的方案。同时，在保证真实度可量化的基础上，成功模拟了社会网络中的echo chamber，power-law、同时可以用于KOL selection。
%% 论文提出了基于LLM 强化学习的social graph generator Graphia。回答了social graph generation中，各个social agent，who/how/when to interact的问题。

%% 不足之处在于，尚未给出why to interact的解决方案。未来的工作可能结合心理学的理论进行llm behavior的alignment量化和分析。

\section*{Acknowledgments}
This research was supported in part by National Natural Science Foundation of China (No. 92470128, No. U2241212), by Alibaba Group through Alibaba Innovative Research Program. We also wish to acknowledge the support provided by the fund for building world-class universities (disciplines) of Renmin University of China, by Engineering Research Center of Next-Generation Intelligent Search and Recommendation, Ministry of Education, by Intelligent Social Governance Interdisciplinary Platform, Major Innovation \& Planning Interdisciplinary Platform for the "Double-First Class" Initiative, Public Policy and Decision-making Research Lab, and Public Computing Cloud, Renmin University of China.

% This document has been adapted
% by Steven Bethard, Ryan Cotterell and Rui Yan
% from the instructions for earlier ACL and NAACL proceedings, including those for
% ACL 2019 by Douwe Kiela and Ivan Vuli\'{c},
% NAACL 2019 by Stephanie Lukin and Alla Roskovskaya,
% ACL 2018 by Shay Cohen, Kevin Gimpel, and Wei Lu,
% NAACL 2018 by Margaret Mitchell and Stephanie Lukin,
% Bib\TeX{} suggestions for (NA)ACL 2017/2018 from Jason Eisner,
% ACL 2017 by Dan Gildea and Min-Yen Kan,
% NAACL 2017 by Margaret Mitchell,
% ACL 2012 by Maggie Li and Michael White,
% ACL 2010 by Jing-Shin Chang and Philipp Koehn,
% ACL 2008 by Johanna D. Moore, Simone Teufel, James Allan, and Sadaoki Furui,
% ACL 2005 by Hwee Tou Ng and Kemal Oflazer,
% ACL 2002 by Eugene Charniak and Dekang Lin,
% and earlier ACL and EACL formats written by several people, including
% John Chen, Henry S. Thompson and Donald Walker.
% Additional elements were taken from the formatting instructions of the \emph{International Joint Conference on Artificial Intelligence} and the \emph{Conference on Computer Vision and Pattern Recognition}.


% \bibliography{custom}
\bibliography{format}

\begin{thebibliography}{39}
\providecommand{\natexlab}[1]{#1}

\bibitem[{Chen et~al.(2023)Chen, He, Han, and Liu}]{EDGE}
Xiaohui Chen, Jiaxing He, Xu~Han, and Liping Liu. 2023.
\newblock Efficient and degree-guided graph generation via discrete diffusion modeling.
\newblock In \emph{Int. Conf. Machin. Learn., ICML}, pages 4585--4610. PMLR.

\bibitem[{Cong et~al.(2023)Cong, Zhang, Kang, Yuan, Wu, Zhou, Tong, and Mahdavi}]{graphmixer}
Weilin Cong, Si~Zhang, Jian Kang, Baichuan Yuan, Hao Wu, Xin Zhou, Hanghang Tong, and Mehrdad Mahdavi. 2023.
\newblock \href {https://openreview.net/forum?id=ayPPc0SyLv1} {Do we really need complicated model architectures for temporal networks?}
\newblock In \emph{Proc. Int. Conf. Learn. Represent., ICLR}.

\bibitem[{Del~Vicario et~al.(2016)Del~Vicario, Bessi, Zollo, Petroni, Scala, Caldarelli, Stanley, and Quattrociocchi}]{del2016spreading}
Michela Del~Vicario, Alessandro Bessi, Fabiana Zollo, Fabio Petroni, Antonio Scala, Guido Caldarelli, H~Eugene Stanley, and Walter Quattrociocchi. 2016.
\newblock The spreading of misinformation online.
\newblock \emph{Proceedings of the national academy of Sciences, PNAS}, 113(3):554--559.

\bibitem[{Du et~al.(2025)Du, Li, Jin, Zhang, Li, and Wang}]{GraphMaster}
Enjun Du, Xunkai Li, Tian Jin, Zhihan Zhang, Rong{-}Hua Li, and Guoren Wang. 2025.
\newblock \href {https://doi.org/10.48550/ARXIV.2504.00711} {Graphmaster: Automated graph synthesis via {LLM} agents in data-limited environments}.
\newblock \emph{CoRR}, abs/2504.00711.

\bibitem[{Dubey et~al.(2024)Dubey, Jauhri, Pandey, Kadian, Al{-}Dahle, Letman, Mathur, Schelten, Yang, Fan, Goyal, Hartshorn, Yang, Mitra, Sravankumar, Korenev, Hinsvark, Rao, Zhang, Rodriguez, Gregerson, Spataru, Rozi{\`{e}}re, Biron, Tang, Chern, Caucheteux, Nayak, Bi, Marra, McConnell, Keller, Touret, Wu, Wong, Ferrer, Nikolaidis, Allonsius, Song, Pintz, Livshits, Esiobu, Choudhary, Mahajan, Garcia{-}Olano, Perino, Hupkes, Lakomkin, AlBadawy, Lobanova, Dinan, Smith, Radenovic, Zhang, Synnaeve, Lee, Anderson, Nail, Mialon, Pang, Cucurell, Nguyen, Korevaar, Xu, Touvron, Zarov, Ibarra, Kloumann, Misra, Evtimov, Copet, Lee, Geffert, Vranes, Park, Mahadeokar, Shah, van~der Linde, Billock, Hong, Lee, Fu, Chi, Huang, Liu, Wang, Yu, Bitton, Spisak, Park, Rocca, Johnstun, Saxe, Jia, Alwala, Upasani, Plawiak, Li, Heafield, Stone, and et~al.}]{dubey2024llama}
Abhimanyu Dubey, Abhinav Jauhri, Abhinav Pandey, Abhishek Kadian, Ahmad Al{-}Dahle, Aiesha Letman, Akhil Mathur, Alan Schelten, Amy Yang, Angela Fan, Anirudh Goyal, Anthony Hartshorn, Aobo Yang, Archi Mitra, Archie Sravankumar, Artem Korenev, Arthur Hinsvark, Arun Rao, Aston Zhang, and 82 others. 2024.
\newblock \href {https://doi.org/10.48550/ARXIV.2407.21783} {The llama 3 herd of models}.
\newblock \emph{CoRR}, abs/2407.21783.

\bibitem[{Eigenschink et~al.(2023)Eigenschink, Reutterer, Vamosi, Vamosi, Sun, and Kalcher}]{ieee_survey}
Peter Eigenschink, Thomas Reutterer, Stefan Vamosi, Ralf Vamosi, Chang Sun, and Klaudius Kalcher. 2023.
\newblock Deep generative models for synthetic data: A survey.
\newblock \emph{IEEE Access}, 11:47304--47320.

\bibitem[{Gao et~al.(2024)Gao, Lan, Li, Yuan, Ding, Zhou, Xu, and Li}]{llm_simulation_survey1}
Chen Gao, Xiaochong Lan, Nian Li, Yuan Yuan, Jingtao Ding, Zhilun Zhou, Fengli Xu, and Yong Li. 2024.
\newblock Large language models empowered agent-based modeling and simulation: A survey and perspectives.
\newblock \emph{Humanities and Social Sciences Communications}, 11(1):1--24.

\bibitem[{Gao et~al.(2023)Gao, Lan, Lu, Mao, Piao, Wang, Jin, and Li}]{S3}
Chen Gao, Xiaochong Lan, Zhihong Lu, Jinzhu Mao, Jinghua Piao, Huandong Wang, Depeng Jin, and Yong Li. 2023.
\newblock \href {https://doi.org/10.48550/ARXIV.2307.14984} {S\({}^{\mbox{3}}\): Social-network simulation system with large language model-empowered agents}.
\newblock \emph{CoRR}, abs/2307.14984.

\bibitem[{Guo et~al.(2025)Guo, Yang, Zhang, Song, Wang, Zhu, Xu, Zhang, Ma, Bi, Zhang, Yu, Wu, Wu, Gou, Shao, Li, Gao, Liu, Xue, Wang, Wu, Feng, Lu, Zhao, Deng, Ruan, Dai, Chen, Ji, Li, Lin, Dai, Luo, Hao, Chen, Li, Zhang, Xu, Ding, Gao, Qu, Li, Guo, Li, Chen, Yuan, Tu, Qiu, Li, Cai, Ni, Liang, Chen, Dong, Hu, You, Gao, Guan, Huang, Yu, Wang, Zhang, Zhao, Wang, Zhang, Xu, Xia, Zhang, Zhang, Tang, Zhou, Li, Wang, Li, Tian, Huang, Zhang, Wang, Chen, Du, Ge, Zhang, Pan, Wang, Chen, Jin, Chen, Lu, Zhou, Chen, Ye, Wang, Yu, Zhou, Pan, Li, Zhou, Wu, Yun, Pei, Sun, Wang, Zeng, Liu, Liang, Gao, Yu, Zhang, Xiao, An, Liu, Wang, Chen, Nie, Cheng, Liu, Xie, Liu, Yang, Li, Su, Lin, Li, Jin, Shen, Chen, Sun, Wang, Song, Zhou, Wang, Shan, Li, Wang, Wei, Zhang, Xu, Li, Zhao, Sun, Wang, Yu, Zhang, Shi, Xiong, He, Piao, Wang, Tan, Ma, Liu, Guo, Ou, Wang, Gong, Zou, He, Xiong, Luo, You, Liu, Zhou, Zhu, Huang, Li, Zheng, Zhu, Ma, Tang, Zha, Yan, Ren, Ren, Sha, Fu, Xu, Xie, Zhang, Hao, Ma, Yan, Wu, Gu, Zhu, Liu, Li, Xie, Song, Pan, Huang, Xu, Zhang, and Zhang}]{guo2025deepseek}
Daya Guo, Dejian Yang, Haowei Zhang, Junxiao Song, Peiyi Wang, Qihao Zhu, Runxin Xu, Ruoyu Zhang, Shirong Ma, Xiao Bi, Xiaokang Zhang, Xingkai Yu, Yu~Wu, Z.~F. Wu, Zhibin Gou, Zhihong Shao, Zhuoshu Li, Ziyi Gao, Aixin Liu, and 175 others. 2025.
\newblock \href {https://doi.org/10.1038/s41586-025-09422-z} {{DeepSeek}-{R1} incentivizes reasoning in {LLMs} through reinforcement learning}.
\newblock \emph{Nature}, 645(8081):633--638.

\bibitem[{Gupta et~al.(2022)Gupta, Manchanda, Bedathur, and Ranu}]{TIGGER}
Shubham Gupta, Sahil Manchanda, Srikanta Bedathur, and Sayan Ranu. 2022.
\newblock \href {https://doi.org/10.1609/AAAI.V36I6.20638} {{TIGGER:} scalable generative modelling for temporal interaction graphs}.
\newblock In \emph{Proc. AAAI Conf. Artif. Intell., AAAI}, pages 6819--6828.

\bibitem[{Hosseini et~al.(2025)Hosseini, Simini, Vishwanath, and Hoffmann}]{DGGen}
Ryien Hosseini, Filippo Simini, Venkatram Vishwanath, and Henry Hoffmann. 2025.
\newblock \href {https://doi.org/10.1609/AAAI.V39I16.33896} {A deep probabilistic framework for continuous time dynamic graph generation}.
\newblock In \emph{Proc. AAAI Conf. Artif. Intell., AAAI}, pages 17249--17257.

\bibitem[{Ji et~al.(2025)Ji, Lei, Bi, Wei, Chen, Lin, Pan, Li, and Ding}]{GAG}
Jiarui Ji, Runlin Lei, Jialing Bi, Zhewei Wei, Xu~Chen, Yankai Lin, Xuchen Pan, Yaliang Li, and Bolin Ding. 2025.
\newblock \href {https://aclanthology.org/2025.findings-acl.78/} {Llm-based multi-agent systems are scalable graph generative models}.
\newblock In \emph{Find. Annu. Meet. Assoc. Comput Linguist., ACL}, pages 1492--1523.

\bibitem[{Lei et~al.(2025)Lei, Ji, Ding, Yi, Wei, Liu, and Hong}]{GAD}
Runlin Lei, Jiarui Ji, Haipeng Ding, Lu~Yi, Zhewei Wei, Yongchao Liu, and Chuntao Hong. 2025.
\newblock \href {https://doi.org/10.48550/ARXIV.2503.03258} {Exploring the potential of large language models as predictors in dynamic text-attributed graphs}.
\newblock \emph{CoRR}, abs/2503.03258.

\bibitem[{Li et~al.(2024)Li, Dong, Chen, Su, Zhou, Ai, Ye, and Liu}]{llm-judge-problem}
Haitao Li, Qian Dong, Junjie Chen, Huixue Su, Yujia Zhou, Qingyao Ai, Ziyi Ye, and Yiqun Liu. 2024.
\newblock \href {https://doi.org/10.48550/ARXIV.2412.05579} {Llms-as-judges: {A} comprehensive survey on llm-based evaluation methods}.
\newblock \emph{CoRR}, abs/2412.05579.

\bibitem[{Liu et~al.(2024)Liu, Chen, Zhang, Gao, Zhang, and Yan}]{FUSE}
Yuhan Liu, Xiuying Chen, Xiaoqing Zhang, Xing Gao, Ji~Zhang, and Rui Yan. 2024.
\newblock \href {https://api.semanticscholar.org/CorpusID:268385134} {From skepticism to acceptance: Simulating the attitude dynamics toward fake news}.
\newblock In \emph{Proc. Int. Joint Conf. Artif. Intell., IJCAI}.

\bibitem[{McPherson et~al.(2001)McPherson, Smith-Lovin, and Cook}]{mcpherson2001birds}
Miller McPherson, Lynn Smith-Lovin, and James~M Cook. 2001.
\newblock Birds of a feather: Homophily in social networks.
\newblock \emph{Annual review of sociology}, 27(1):415--444.

\bibitem[{Mou et~al.(2024)Mou, Wei, and Huang}]{SoMoSiMuBench}
Xinyi Mou, Zhongyu Wei, and Xuanjing Huang. 2024.
\newblock \href {https://doi.org/10.18653/V1/2024.FINDINGS-ACL.285} {Unveiling the truth and facilitating change: Towards agent-based large-scale social movement simulation}.
\newblock In \emph{Find. Annu. Meet. Assoc. Comput Linguist., ACL}, pages 4789--4809.

\bibitem[{Peng et~al.(2025)Peng, Ji, Lei, Wei, Liu, and Hong}]{peng2025gdgb}
Jie Peng, Jiarui Ji, Runlin Lei, Zhewei Wei, Yongchao Liu, and Chuntao Hong. 2025.
\newblock \href {https://doi.org/10.48550/ARXIV.2507.03267} {{GDGB:} {A} benchmark for generative dynamic text-attributed graph learning}.
\newblock \emph{CoRR}, abs/2507.03267.

\bibitem[{Piao et~al.(2025)Piao, Yan, Zhang, Li, Yan, Lan, Lu, Zheng, Wang, Zhou, Gao, Xu, Zhang, Rong, Su, and Li}]{piao2025agentsociety}
Jinghua Piao, Yuwei Yan, Jun Zhang, Nian Li, Junbo Yan, Xiaochong Lan, Zhihong Lu, Zhiheng Zheng, Jing~Yi Wang, Di~Zhou, Chen Gao, Fengli Xu, Fang Zhang, Ke~Rong, Jun Su, and Yong Li. 2025.
\newblock \href {https://doi.org/10.48550/ARXIV.2502.08691} {Agentsociety: Large-scale simulation of llm-driven generative agents advances understanding of human behaviors and society}.
\newblock \emph{CoRR}, abs/2502.08691.

\bibitem[{Rossetti et~al.(2024)Rossetti, Stella, Cazabet, Abramski, Cau, Citraro, Failla, Improta, Morini, and Pansanella}]{Y_social}
Giulio Rossetti, Massimo Stella, R{\'{e}}my Cazabet, Katherine Abramski, Erica Cau, Salvatore Citraro, Andrea Failla, Riccardo Improta, Virginia Morini, and Valentina Pansanella. 2024.
\newblock \href {https://doi.org/10.48550/ARXIV.2408.00818} {Y social: an llm-powered social media digital twin}.
\newblock \emph{CoRR}, abs/2408.00818.

\bibitem[{Rossi et~al.(2020)Rossi, Chamberlain, Frasca, Eynard, Monti, and Bronstein}]{TGN}
Emanuele Rossi, Ben Chamberlain, Fabrizio Frasca, Davide Eynard, Federico Monti, and Michael Bronstein. 2020.
\newblock Temporal graph networks for deep learning on dynamic graphs.
\newblock In \emph{ICML 2020 Workshop on Graph Representation Learning}.

\bibitem[{Sakurai et~al.(2025)Sakurai, Togo, Ogawa, and Haseyama}]{LIKR}
Keigo Sakurai, Ren Togo, Takahiro Ogawa, and Miki Haseyama. 2025.
\newblock \href {https://doi.org/10.1007/978-3-031-88711-6\_17} {{LLM} is knowledge graph reasoner: Llm's intuition-aware knowledge graph reasoning for cold-start sequential recommendation}.
\newblock In \emph{Proc. Eur. Conf. Inf. Retr., ECIR}, volume 15573 of \emph{Lecture Notes in Computer Science}, pages 263--278.

\bibitem[{Shao et~al.(2024)Shao, Wang, Zhu, Xu, Song, Bi, Zhang, Zhang, Li et~al.}]{grpo}
Zhihong Shao, Peiyi Wang, Qihao Zhu, Runxin Xu, Junxiao Song, Xiao Bi, Haowei Zhang, Mingchuan Zhang, YK~Li, and 1 others. 2024.
\newblock Deepseekmath: Pushing the limits of mathematical reasoning in open language models.
\newblock \emph{arXiv preprint arXiv:2402.03300}.

\bibitem[{Wang et~al.(2025{\natexlab{a}})Wang, Liu, Yang, and Chen}]{coling_echo_chamber}
Chenxi Wang, Zongfang Liu, Dequan Yang, and Xiuying Chen. 2025{\natexlab{a}}.
\newblock \href {https://aclanthology.org/2025.coling-main.264/} {Decoding echo chambers: Llm-powered simulations revealing polarization in social networks}.
\newblock In \emph{Proc. Int. Conf. Comput. Linguist., COLING}, pages 3913--3923.

\bibitem[{Wang et~al.(2025{\natexlab{b}})Wang, Wang, Zhang, Yuan, Xu, Huang, Yuan, Guo, Chen, Zhou et~al.}]{wangcoser}
Xintao Wang, Heng Wang, Yifei Zhang, Xinfeng Yuan, Rui Xu, Jen-tse Huang, Siyu Yuan, Haoran Guo, Jiangjie Chen, Shuchang Zhou, and 1 others. 2025{\natexlab{b}}.
\newblock \href {https://openreview.net/forum?id=BOrR7YqKUt} {Co{SER}: Coordinating {LLM}-based persona simulation of established roles}.
\newblock In \emph{Proc. Int. Conf. Machin. Learn., ICML}.

\bibitem[{Wang et~al.(2021)Wang, Chang, Liu, Leskovec, and Li}]{CAWN}
Yanbang Wang, Yen{-}Yu Chang, Yunyu Liu, Jure Leskovec, and Pan Li. 2021.
\newblock \href {https://openreview.net/forum?id=KYPz4YsCPj} {Inductive representation learning in temporal networks via causal anonymous walks}.
\newblock In \emph{Proc. Int. Conf. Learn. Represent., ICLR}. OpenReview.net.

\bibitem[{Yang et~al.(2025)Yang, Li, Yang, Zhang, Hui, Zheng, Yu, Gao, Huang, Lv, Zheng, Liu, Zhou, Huang, Hu, Ge, Wei, Lin, Tang, Yang, Tu, Zhang, Yang, Yang, Zhou, Lin, Dang, Bao, Yang, Yu, Deng, Li, Xue, Li, Zhang, Wang, Zhu, Men, Gao, Liu, Luo, Li, Tang, Yin, Ren, Wang, Zhang, Ren, Fan, Su, Zhang, Zhang, Wan, Liu, Wang, Cui, Zhang, Zhou, and Qiu}]{yang2025qwen3}
An~Yang, Anfeng Li, Baosong Yang, Beichen Zhang, Binyuan Hui, Bo~Zheng, Bowen Yu, Chang Gao, Chengen Huang, Chenxu Lv, Chujie Zheng, Dayiheng Liu, Fan Zhou, Fei Huang, Feng Hu, Hao Ge, Haoran Wei, Huan Lin, Jialong Tang, and 40 others. 2025.
\newblock \href {https://doi.org/10.48550/ARXIV.2505.09388} {Qwen3 technical report}.
\newblock \emph{CoRR}, abs/2505.09388.

\bibitem[{Yang et~al.(2024)Yang, Yang, Hui, Zheng, Yu, Zhou, Li, Li, Liu, Huang, Dong, Wei, Lin, Tang, Wang, Yang, Tu, Zhang, Ma, Yang, Xu, Zhou, Bai, He, Lin, Dang, Lu, Chen, Yang, Li, Xue, Ni, Zhang, Wang, Peng, Men, Gao, Lin, Wang, Bai, Tan, Zhu, Li, Liu, Ge, Deng, Zhou, Ren, Zhang, Wei, Ren, Liu, Fan, Yao, Zhang, Wan, Chu, Liu, Cui, Zhang, Guo, and Fan}]{yang2024qwen2technicalreport}
An~Yang, Baosong Yang, Binyuan Hui, Bo~Zheng, Bowen Yu, Chang Zhou, Chengpeng Li, Chengyuan Li, Dayiheng Liu, Fei Huang, Guanting Dong, Haoran Wei, Huan Lin, Jialong Tang, Jialin Wang, Jian Yang, Jianhong Tu, Jianwei Zhang, Jianxin Ma, and 43 others. 2024.
\newblock \href {https://arxiv.org/abs/2407.10671} {Qwen2 technical report}.
\newblock \emph{Preprint}, arXiv:2407.10671.

\bibitem[{Yao et~al.(2025)Yao, Zhang, Ou, Zuo, Yang, and Dong}]{yao2025social}
Junchi Yao, Hongjie Zhang, Jie Ou, Dingyi Zuo, Zheng Yang, and Zhicheng Dong. 2025.
\newblock Social opinions prediction utilizes fusing dynamics equation with llm-based agents.
\newblock \emph{Scientific Reports}, 15(1):15472.

\bibitem[{Yu et~al.(2025)Yu, Qi, Zhao, Nottingham, Xuan, Majumder, Zhu, Liang, and You}]{SOTOPIA-RL}
Haofei Yu, Zhengyang Qi, Yining Zhao, Kolby Nottingham, Keyang Xuan, Bodhisattwa~Prasad Majumder, Hao Zhu, Paul~Pu Liang, and Jiaxuan You. 2025.
\newblock \href {https://doi.org/10.48550/ARXIV.2508.03905} {Sotopia-rl: Reward design for social intelligence}.
\newblock \emph{CoRR}, abs/2508.03905.

\bibitem[{Yu et~al.(2023)Yu, Sun, Du, and Lv}]{Dygformer}
Le~Yu, Leilei Sun, Bowen Du, and Weifeng Lv. 2023.
\newblock \href {http://papers.nips.cc/paper\_files/paper/2023/hash/d611019afba70d547bd595e8a4158f55-Abstract-Conference.html} {Towards better dynamic graph learning: New architecture and unified library}.
\newblock In \emph{Proc. Adv. neural inf. proces. syst., NeurIPS}.

\bibitem[{Zhang et~al.(2024{\natexlab{a}})Zhang, Chen, Yang, Feng, Liang, Shao, and Ying}]{DTGB}
Jiasheng Zhang, Jialin Chen, Menglin Yang, Aosong Feng, Shuang Liang, Jie Shao, and Rex Ying. 2024{\natexlab{a}}.
\newblock \href {http://papers.nips.cc/paper\_files/paper/2024/hash/a65d054a407f94c34ecfb598fb540a0d-Abstract-Datasets\_and\_Benchmarks\_Track.html} {{DTGB:} {A} comprehensive benchmark for dynamic text-attributed graphs}.
\newblock In \emph{Proc. Adv. neural inf. proces. syst., NeurIPS}.

\bibitem[{Zhang et~al.(2020)Zhang, Kishore, Wu, Weinberger, and Artzi}]{bertscore}
Tianyi Zhang, Varsha Kishore, Felix Wu, Kilian~Q. Weinberger, and Yoav Artzi. 2020.
\newblock \href {https://openreview.net/forum?id=SkeHuCVFDr} {Bertscore: Evaluating text generation with {BERT}}.
\newblock In \emph{Proc. Int. Conf. Learn. Represent., ICLR}.

\bibitem[{Zhang et~al.(2024{\natexlab{b}})Zhang, Chen, Liu, Wang, Hu, and Yan}]{TIS}
Xiaoqing Zhang, Xiuying Chen, Yuhan Liu, Jianzhou Wang, Zhenxing Hu, and Rui Yan. 2024{\natexlab{b}}.
\newblock \href {https://doi.org/10.48550/ARXIV.2411.01143} {A large-scale time-aware agents simulation for influencer selection in digital advertising campaigns}.
\newblock \emph{CoRR}, abs/2411.01143.

\bibitem[{Zhang et~al.(2025)Zhang, Liu, Wang, Hu, Chen, and Yan}]{sagraph_bench}
Xiaoqing Zhang, Yuhan Liu, Jianzhou Wang, Zhenxing Hu, Xiuying Chen, and Rui Yan. 2025.
\newblock \href {https://doi.org/10.1145/3726302.3730334} {Sagraph: {A} large-scale social graph dataset with comprehensive context for influencer selection in marketing}.
\newblock In \emph{Proc. Int. ACM SIGIR Conf. Res. Dev. Inf. Retr., SIGIR}, pages 3733--3742.

\bibitem[{Zheng and Tang(2024)}]{zheng2024simulating}
Wenzhen Zheng and Xijin Tang. 2024.
\newblock Simulating social network with llm agents: An analysis of information propagation and echo chambers.
\newblock In \emph{Proc. Int. Conf. Learn. Represent., ICLR}, pages 63--77. Springer.

\bibitem[{Zhou et~al.(2021)Zhou, Zhang, Peng, Zhang, Li, Xiong, and Zhang}]{zhou2021informer}
Haoyi Zhou, Shanghang Zhang, Jieqi Peng, Shuai Zhang, Jianxin Li, Hui Xiong, and Wancai Zhang. 2021.
\newblock Informer: Beyond efficient transformer for long sequence time-series forecasting.
\newblock In \emph{Proc. AAAI Conf. Artif. Intell., AAAI}, volume~35, pages 11106--11115.

\bibitem[{Zhou et~al.(2025)Zhou, Zhang, Gao, Jiang, and Wang}]{zhou2025personaeval}
Lingfeng Zhou, Jialing Zhang, Jin Gao, Mohan Jiang, and Dequan Wang. 2025.
\newblock \href {https://openreview.net/forum?id=drdrFhKYjP} {Personaeval: Are {LLM} evaluators human enough to judge role-play?}
\newblock In \emph{Proc. Second Conf. Language Modeling, COLM}.

\bibitem[{Zhou et~al.(2024)Zhou, Zhu, Mathur, Zhang, Yu, Qi, Morency, Bisk, Fried, Neubig, and Sap}]{zhousotopia}
Xuhui Zhou, Hao Zhu, Leena Mathur, Ruohong Zhang, Haofei Yu, Zhengyang Qi, Louis{-}Philippe Morency, Yonatan Bisk, Daniel Fried, Graham Neubig, and Maarten Sap. 2024.
\newblock \href {https://openreview.net/forum?id=mM7VurbA4r} {{SOTOPIA:} interactive evaluation for social intelligence in language agents}.
\newblock In \emph{Proc. Int. Conf. Learn. Represent., ICLR}.

\end{thebibliography}

\appendix
% 可选：将“Appendix Contents”加入主目录
\addcontentsline{toc}{section}{Appendix: Contents}
% \section*{Appendix: Contents}

% ✅ 关键步骤：开启名为 "appendices" 的局部目录
\startcontents[appendices]
\addcontentsline{toc}{section}{Appendix: Contents} % 可选：是否加入主 ToC

% 生成仅包含附录章节的局部目录
\printcontents[appendices]{}{1}{}

\newpage

\section{Details of Dataset}
\label{app:dataset_details}
We provide additional details about the three social network datasets used in our experiments.

\begin{itemize}
    \item \textbf{Propagate-En.} A product-sharing social network collected from an e-commerce platform, where nodes represent taokes (content influencers) and edges indicate forwarding behaviors of promotional content.
    \item \textbf{Weibo Tech.} A subgraph of the Weibo network focused on technology-related topics, capturing information diffusion among tech influencers.
    \item \textbf{Weibo Daily.} A general-topic Weibo network with broader user coverage, reflecting daily social interactions and news propagation.
\end{itemize}

All data are binned into daily snapshots. For temporal splitting, let $ T_2 $ denote the total duration (in days). We define $ \tau = \lfloor 0.15 \times T_2 \rfloor $ and use:
\begin{itemize}
    \item \textbf{Train Split.} $[0, T)$, where $T = T_2 - 3\tau$
    \item \textbf{Validation Split.} $[T, T + \tau)$, used to predict $[T + \tau, T + 2\tau)$
    \item \textbf{Test Split}: $[T + \tau, T + 2\tau)$, used to predict $[T + 2\tau, T_2)$
\end{itemize}
Thus, both the training, validation, and test splits use an input history of length $T$ days to predict the next $\tau$ days, ensuring a consistent evaluation protocol across all phases.
The full statistics, including average edge duration (i.e., the mean lifespan of each edge in days), are shown in Table~\ref{tab:appendix_dataset}.

\begin{table*}[h]
\centering
\caption{Extended dataset statistics.}
\label{tab:appendix_dataset}
  \begin{adjustbox}{width=.65\textwidth, totalheight=.65\textheight, keepaspectratio}
\begin{tabular}{lrrr}
\toprule
\textbf{Dataset} & \textbf{Propagate-En} & \textbf{Weibo Tech} & \textbf{Weibo Daily} \\
\midrule
Node Count                    & 5,634     & 20,768    & 66,501     \\
Edge Count                    & 16,962    & 58,930    & 195,202    \\
Input Length (Days)           & 18        & 5         & 19         \\
Prediction Length (Days)     & 4         & 1         & 4          \\
Total Length (Days)       & 30        & 8         & 31         \\
Avg. Edge (Days)     & 565.40    & 7,366.25  & 6,296.84   \\
Number of Categories          & 534       & 2         & 2          \\
Input Edge Count (Test)            & 10,119    & 45,623    & 125,073    \\
Prediction Edge Count (Test)        & 2,283     & 8,618     & 47,980     \\
\bottomrule
\end{tabular}
\end{adjustbox}
\end{table*}

\section{Details of Metric}

We provide detailed mathematical formulations and implementation specifics for the IDGG and TDGG social fidelity Scores introduced in Section~\ref{sec:experiment_setup}.

First, we define the dataset-wise normalization function for different metrics. To map all component metrics to [0,1] with a positive direction (higher is better), we apply min–max normalization per metric within each dataset. 

For positive-direction metric value $x$, its normalized score is:
\begin{equation*}
    \overline{x} = \frac{x - \min(x)}{\max(x) - \min(x)},
\end{equation*}
where $\min(x)$ and $\max(x)$ are computed over the metric values of across all models evaluated on the same dataset. 
The positive-direction metrics include: $\mathrm{R}@100\text{-Easy}$, $\mathrm{R}@100\text{-Hard}$, $\mathrm{R}@100\text{-All}$, $\mathrm{ACC}$, ROUGE-L, BERTScore-F1, EO, and $\mathrm{P}@100\text{-KOL}$.

For negative-direction metrics (smaller is better) $x$, we use reversed normalized score as:
\begin{equation*}
    1 - \overline{x} = 1 - \frac{x - \min(x)}{\max(x) - \min(x)}.
\end{equation*}
The negative-direction metrics include: $\mathrm{MMD.D}^2, \mathrm{MMD.C}^2,\mathrm{MMD.S}^2$, D, and Chambers Diff.
This normalization is applied independently to each metric prior to aggregation into the final scores.

\subsection{TDGG Social Fidelity Score}
\label{app:tdgg_details}

The TDGG social fidelity score ($S_{\text{TDGG}}$) evaluates micro-level behavioral authenticity by combining destination selection fidelity and edge generation quality. 

\noindent\textbf{Destination Selection.} \ Following PersonaEval, we assess whether models select accurate partners using $\mathrm{R}@100$~\cite{zhou2025personaeval}, the fraction of true destinations ranked among the top-100 predicted nodes. To differentiate difficulty, we categorize target nodes by out-degree $d(u)$: \textbf{Easy}: $d(u) > \mathrm{70}\text{-th percentile degree}$ (i.e., hub nodes), \textbf{Hard}: $d(u) \leq \mathrm{70}\text{-th percentile degree}$ (i.e., non-hub nodes). We compute 
These are normalized and averaged into the selection score:
\begin{align*}
S_{\text{selection}} & = \frac{1}{3}(
\overline{\mathrm{R}@100\text{-Easy}} + \\
& \overline{\mathrm{R}@100\text{-Hard}} + \overline{\mathrm{R}@100\text{-All}} 
).  
\end{align*}

\noindent\textbf{Edge Generation.} \ To evaluate the quality of generated interaction messages, we assess both the predicted message category $\hat{y}_{u\to v}$ against the reference label $y_{u\to v}$, and the gener ated message text $\hat{m}_{u\to v}$ against the reference message $m_{u\to v}$ using ROUGE-L and BERTScore-F1.

\begin{itemize}
    \item \textbf{Category Accuracy.} Measures the accuracy ($\mathrm{ACC}$) of predicted message types (e.g., question, greeting, request).For the set of evaluated edges $\mathcal{E}$, it is defined as::
    $$
    \mathrm{ACC}=\frac{1}{|\mathcal{E}|} \sum_{({u, v}) \in \mathcal{E}} \mathbb{I}\left(\hat{y}_{u\to v}=y_{u\to v}\right),
    $$ where $\hat{y}_e$ and $y_e$ denote the predicted and true message categories for edge $e$, respectively, and $\mathbb{I}(\cdot)$ is the indicator function.

    \item \textbf{ROUGE-L.} Evaluates the similarity between generated and reference edge messages using the longest common subsequence (LCS). It measures n-gram co-occurrence with flexibility in word order, making it robust to syntactic variations. Formally:
    $$
    \text{ROUGE-L} = \frac{
        \sum_{e \in \mathcal{E}} \text{LCS}(\hat{m}_e, m_e)
    }{
        \sum_{e \in \mathcal{E}} |m_e|
    },
    $$
    where $\text{LCS}(\hat{m}_e, m_e)$ denotes the length of the longest common subsequence between the generated message $\hat{m}_e$ and its ground truth $m_e$, and $|m_e|$ is the token length of the reference message. The final score is computed as an average over all edges in the evaluation set.

    \item \textbf{BERTScore-F1.} We adopt BERTScore to compute a contextual F1 score between generated and reference edge messages~\cite{bertscore}, leveraging pretrained contextual embeddings for more semantically meaningful similarity measurement. Formally, for each edge $ e \in \mathcal{E} $, we compare the generated message $\hat{m}_e$ with its ground truth $m_e$, and compute BERTScore over the entire set of ($m_e,\hat{m}_e$) pairs.

\end{itemize}

These metrics are normalized and averaged into the edge sub-score:
$$
S_{\text{edge}}  = \frac{1}{3}( \overline{\mathrm{ACC}} + \overline{\text{ROUGE-L}} + \overline{\text{BERTScore-F1}} ).  $$

\noindent \textbf{(3) TDGG Social Fidelity Score.} \ The final score is computed as:
$$
S_{\text{TDGG}} = 0.5 \cdot S_{\text{selection}} + 0.5 \cdot S_{\text{edge}},
$$
with equal weighting reflecting balanced importance between structural attention and semantic realism.

\subsection{IDGG Social Fidelity Score}
\label{app:idgg_details}

The IDGG social fidelity score ($S_{\text{IDGG}}$) evaluates macro-level realism through macro-level structural and phenomenological realism.

\noindent\textbf{Macro Structure Fidelity.} \ We assess structural similarity using distributional distances measured by Maximum Mean Discrepancy (MMD) with an RBF kernel:
\begin{align*}
\mathrm{MMD}^2(X, Y) & = \frac{1}{n^2} \sum_{i,j=1}^n k(x_i, x_j) \\
& + \frac{1}{m^2} \sum_{i,j=1}^m k(y_i, y_j) \\
& - \frac{2}{nm} \sum_{i=1}^n \sum_{j=1}^m k(x_i, y_j),  
\end{align*}
where $k(\cdot,\cdot)$ is the RBF kernel $k(a,b)=\exp(-\|a-b\|^2 / 2v^2)$.
In our evaluation, we report $\mathrm{MMD.D}^2$ for degree distribution, $\mathrm{MMD.C}^2$ for cluster coefficient, and $\mathrm{MMD.S}^2$ for spectral properties, consistent with common practice in graph generation literature.
Additionally, we compute edge overlap ratio for future edges:
$$
\mathrm{EO} = \frac{|\hat{\mathcal{E}}_{\text{fut}} \cap \mathcal{E}_{\text{fut}}|}{|\mathcal{E}_{\text{fut}}|}.
$$
These metrics are normalized and averaged into the structure fidelity score:
\begin{align*}
S_{\text{structure}} & = \frac{1}{4}((1 - \overline{\mathrm{MMD.D}^2}) + \\
& (1 - \overline{\mathrm{MMD.C}^2}) + \\
& (1 - \overline{\mathrm{MMD.S}^2}) +
\overline{\text{EO}}
).  
\end{align*}
.

\noindent\textbf{Macro Phenomenon Consistency.} \ We evaluate three canonical social phenomena:

\begin{itemize}
  \item \textbf{Influencer Selection.} 
  Inspired by SaGraph~\cite{sagraph_bench}, we evaluate the model's ability to predict key opinion leaders (KOLs) using $\mathrm{P}@100\text{-KOL}$, which measures the precision in recovering the most influential nodes in the future graph.
  We define the ground-truth KOLs as the 100 nodes with the highest PageRank scores in the reference future graph $\mathcal{G}_{\text{fut}}$. 
  The predicted KOLs are taken as the top-100 nodes ranked by PageRank in the generated graph $\hat{\mathcal{G}}_{\text{fut}}$.
  $\mathrm{P}@100\text{-KOL}$ is then computed as the fraction of true KOLs that appear among the predicted top-100. A higher score indicates better alignment in capturing central influencers.

  \item \textbf{Echo Chamber Alignment.} 
  To quantify ideological polarization, we detect tightly-knit, ideologically homogeneous communities (i.e., echo chambers) in both the reference graph $\mathcal{G}_{\text{fut}}$ and the generated graph $\hat{\mathcal{G}}_{\text{fut}}$, following prior work~\cite{zheng2024simulating,coling_echo_chamber}. 
  The number of such chambers is extracted via community detection under polarization constraints.
  We then measure the deviation in chamber count:
  \begin{align*}
      \Delta C = \left| |\mathrm{Chambers}(\mathcal{G}_{\text{fut}})| - |\mathrm{Chambers}(\hat{\mathcal{G}}_{\text{fut}})| \right|.
  \end{align*}
  A smaller $\Delta C$ indicates better preservation of macro-level social fragmentation patterns.
  \item \textbf{Power-law Degree Distribution.} 
  Following GDGB~\cite{peng2025gdgb}, we assess how well the generated graph preserves the heavy-tailed nature of real-world networks.
  We fit a power-law distribution $p(k) \sim k^{-\alpha}$ to the degree sequence of the reference graph using maximum likelihood estimation, with $x_{\min} = 2$ for all datasets.
  The goodness-of-fit is evaluated via the Kolmogorov-Smirnov (KS) distance between the empirical and fitted distributions.
  Additionally, we compute the deviation in the estimated power-law exponent:
  \begin{align*}
      \Delta \alpha = |\alpha_{\text{ref}} - \alpha_{\text{gen}}|,
  \end{align*}
  where $\alpha_{\text{ref}}$ and $\alpha_{\text{gen}}$ are the exponents fitted on $\mathcal{G}_{\text{fut}}$ and $\hat{\mathcal{G}}_{\text{fut}}$, respectively. Smaller $\Delta \alpha$ indicates better reproduction of scale-free characteristics.

\end{itemize}

These metrics are normalized and averaged into the phenomenon replication score:
\begin{align*}
S_{\text{phenomenon}} & = \frac{1}{3}(
\overline{\mathrm{P}@100\text{-KOL}} + \\
& (1 - \overline{\Delta C}) + (1 - \overline{\Delta \alpha}) 
).  
\end{align*}

\noindent\textbf{IDGG Social Fidelity Score.} \ The final score is computed as:
$$
S_{\text{IDGG}} = 0.5 \cdot S_{\text{structure}} + 0.5 \cdot S_{\text{phenomenon}},
$$
with equal weighting reflecting balanced importance between macro structure alignment and macro phenomenon replication.

\subsection{Degree Prediction Metrics}
\label{app:degree_metrics}

To evaluate the Activity-Predictor in source node degree prediction, we adopt three distributional discrepancy metrics based on histogram comparisons between predicted and ground-truth out-degree distributions. Let $d(v)$ denote the true out-degree of node $v$, $\hat{d}(v)$ its predicted value, and $V$ the set of all nodes. We first construct empirical histograms over predefined bins:
\begin{align*}
    \hat{p}_i & = \frac{\left|\{v \in V : d(v) \in \text{bin}_i\}\right|}{|V|}, \\
    \hat{q}_i & = \frac{\left|\{v \in V : \hat{d}(v) \in \text{bin}_i\}\right|}{|V|},
\end{align*}
where $\hat{p}$ and $\hat{q}$ represent the normalized frequency distributions of true and predicted degrees, respectively.
We then compute the following metrics:

\begin{itemize}
    \item \textbf{Wasserstein Distance.} We measure the Wasserstein distance between the cumulative distribution functions (CDFs) derived from $\hat{p}$ and $\hat{q}$. The CDFs are defined as:
  \begin{align*}
      \hat{P}_i = \sum_{j=1}^{i} \hat{p}_j, \quad \hat{Q}_i = \sum_{j=1}^{i} \hat{q}_j.
  \end{align*}
  The Wasserstein distance is then given by:
  \begin{align*}
      W = \frac{1}{n} \sum_{i=1}^{n} \left| \hat{P}_i - \hat{Q}_i \right|,
  \end{align*}
  where $n$ is the number of bins. The Wasserstein distance quantifies the minimum total cost required to transform the distribution $\hat{p}$ into $\hat{q}$, interpreting bin differences as transportation distances. It is sensitive to shifts in distributional location and shape, making it well-suited for comparing degree distributions.

  \item \textbf{KL-Divergence.} We measure the relative entropy from $\hat{q}$ to $\hat{p}$ using:
  \begin{align*}
      D_{\mathrm{KL}}(\hat{p} \parallel \hat{q}) = \sum_{i=1}^{n} \hat{p}_i \log\left( \frac{\hat{p}_i + \epsilon}{\hat{q}_i + \epsilon} \right),
  \end{align*}
  where $\epsilon = 10^{-10}$ is a small constant added to prevent numerical instability due to zero probabilities. KL divergence quantifies how much information is lost when $\hat{q}$ is used to approximate $\hat{p}$, with lower values indicating better alignment.

  \item \textbf{MMD.OD.} We measure the distribution discrepancy between predicted and true out-degree distributions using the Maximum Mean Discrepancy (MMD) with a Gaussian RBF kernel. While we report $\mathrm{MMD}^2$ in the main IDGG task, for degree prediction we take the square root to obtain a more interpretable scale:
  \begin{align*}
      \mathrm{MMD.OD} = \sqrt{\mathrm{MMD}^2(\hat{p}, \hat{q})}.
  \end{align*}
  This ensures that $\mathrm{MMD.OD}$ has the same units as node degrees, providing an intuitive estimate of distributional divergence in terms of average activity level mismatch.

\end{itemize}

\section{Implementation Details}
\begin{table*}[h]
\centering
\caption{Training Hyperparameter Configuration for Graphia RL Components}
\label{tab:rl_params}
\begin{adjustbox}{width=\textwidth, totalheight=.8\textheight, keepaspectratio}
\begin{tabular}{lllllllllll}
\toprule
Component & Step & $K_{1}$ & $\alpha$ & Component & Step & $\beta_{\text{min}}$ & $\beta_{\text{max}}$ & GNN Rewarder & LLM Rewarder &  Interleave Ratio\\
 &&  &  & & &  & &  & & (GNN:LLM)\\
\midrule
\multicolumn{11}{c}{\textbf{Propagate-En}} \\
\midrule
Graphia-Q & 300 & 3$K_{2}$ & 5 & Graphia-E & 100 & 1 & 5 & GraphMixer & Qwen3-8B & 4:1\\
\midrule
\multicolumn{11}{c}{\textbf{Weibo-Tech}} \\
\midrule
Graphia-Q & 50 & 100 & 1 & Graphia-E & 100 & 1 & 5 & GraphMixer & Qwen3-8B & 1:1\\
\midrule
\multicolumn{11}{c}{\textbf{Weibo-Daily}} \\
\midrule
Graphia-Q & 100 & 1000 & 5 & Graphia-E & 100 & 1 & 5 & GraphMixer & Qwen3-8B & 1:1\\
\bottomrule
\end{tabular}
\end{adjustbox}
\end{table*}
\subsection{Implementation of Graphia}
\label{app:destination_selection}
%% to be done, Graphia-Q和Graphia-E的训练参数、训练数据格式等等。
\noindent\textbf{Destination Selection.} \ For each source node $u$ at time $t$, the goal is to select $K_2 = \mathrm{round}\big( \hat{d}_t(u) \big)$ destination nodes based on a query and behavior filter generated by Graphia-Q. The Graphia-Q generates a textual query to constrain the search space.
First, we sample fixed constant number $K_1$ of candidate destination nodes in two epochs.
\begin{itemize}
    \item First, a textual query is generated to describe the desired characteristics of the target (e.g., ``a user interested in fitness gear''). To reflect real-world social dynamics, the system first retrieves top-matching nodes from $u$'s historical neighbors. The candidate node are ranked based on this query and historical neighbor nodes using cosine similarity of BERT embeddings.  
    \item Second, if fewer than $K_1$ valid neighbors are available, the system expands the search to the general population of profiles. We retrieve and rank nodes from historical neighbors of $u$  using the textual query, scoring via cosine similarity of BERT embeddings.
\end{itemize}

Then, all candidate lists are merged in order of priority: neighbors first, then filtered general nodes. Duplicates are removed while preserving ranking order, and the final list is truncated to $K_2$ destination nodes.

\noindent\textbf{Edge Generation.} \
We employ domain interleaved sampling with a fixed 4:1 ratio (category:message) for Propagate-En and a 1:1 ratio (category:message) for Weibo Tech and Weibo Daily. For each domain, we define a task-specific reward function while enforcing a shared output format to ensure structural consistency during generation.
To train the GraphMixer as a reward model, we follow the training protocol of DTGB~\cite{DTGB}. The detailed training configuration is summarized in Table~\ref{tab:params_graphmixer}.

\begin{table}[h]
\centering
\caption{Training Hyperparameter Configuration for GraphMixer}
\label{tab:params_graphmixer}
\small
\begin{tabular}{ll}
\toprule
Parameter Type & Configuration \\
\midrule
\multicolumn{2}{c}{\textbf{Model Architecture}} \\
Number of GNN Layers & 2 \\
Dropout Rate & 0.1 \\
\midrule
\multicolumn{2}{c}{\textbf{Sampling Strategy}} \\
Number of Neighbors & 20 \\
Sampling Method & Recent \\
\midrule
\multicolumn{2}{c}{\textbf{Training Parameters}} \\
Batch Size & 2048 \\
Patience & 5 \\
\bottomrule
\end{tabular}
\end{table}

\noindent\textbf{Training Details.} \ Our training pipeline consists of two stages: supervised fine-tuning (SFT) followed by task-specific reinforcement learning via GRPO~\cite{grpo}.
For destination selection, the input is $\mathcal{M}_{t}(u) + p_u$ (interaction history and source node profile), and the target output is the ground-truth destination node profile $p_v$. For edge generation, the input is $\mathcal{M}_{t}(u,v) + p_u + p_v$, and the model is trained to generate the actual edge message $m_{u \to v}$. In the SFT stage, we perform full-parameter fine-tuning for edge generation. In the SFT stage, we fully fine-tune the edge generation model, but skip SFT stage for destination selection due to negligible gains observed in abalation experiments.
In the RL stage, we optimize each task separately using GRPO with reward shaping based on domain-specific metrics. Training proceeds until convergence, with early stopping triggered by performance on a validation set.  
In the RL stage, we optimize each task separately using GRPO with reward shaping based on domain-specific metrics. Training proceeds until convergence, with early stopping determined by performance on a validation set.  
To ensure the validation set reflects a meaningful range of task difficulty, we adopt a difficulty-aware sampling strategy based on the out-degree of source nodes. Specifically, we stratify source nodes into three difficulty tiers—low (1), medium (2), and high (3)—using the 30th and 70th percentiles of the out-degree distribution. Only nodes in the low and medium difficulty tiers (levels 1 and 2) are included in the validation set.
During RL optimization, we monitored the validation accuracy and applied early stopping once the performance plateaued. Detailed training hyperparameters for Graphia-Q and Graphia-E are summarized in Table~\ref{tab:rl_params}.
% As a result, the final RL training steps for Graphia-Q on the three datasets are: 200 steps for Propagate-EN and 50 steps for Weibo-Tech and Weibo-Daily; the final RL training steps for Graphia-E on the three datasets are: 200 steps for Propagate-EN and 100 steps for Weibo-Tech and Weibo-Daily. 

%%
% We set $K_1 = 100$ for Propagate-EN and Weibo-Tech. For Weibo-Daily, due to its significantly larger node count, a smaller $K_1$ would result in sparse or missing rewards; thus, we increase $K_1$ to 1000 to ensure sufficient retrieval coverage and stable reward signals.
% We use Qwen3-8B~\footnote{\url{https://huggingface.co/Qwen/Qwen3-8B}} as the backbone LLM for both \textsc{Graphia-Q} and \textsc{Graphia-E}. The number of training epochs per dataset is as follows: 200 epochs for \textsc{Graphia-Q} on Propagate-En, 50 epochs on Weibo Daily, and 50 epochs on Weibo Tech; 200 epochs for \textsc{Graphia-E} on Propagate-En, 100 epochs on Weibo Daily, and 100 epochs on Weibo Tech. Training and SFT are conducted exclusively on the training set, and early stopping is performed based on validation set performance.
%%

\subsection{Implementation of Baselines}
\label{app:baselines}

To systematically evaluate the role of graph-structured inputs in enhancing LLM-based social simulation, we introduce a \textit{sequentialized} data format. In sequentialized data, all higher-order graph structures (e.g., multi-hop neighborhoods, global topology) are removed; each node's context is represented solely as a flat sequence of its one-hop neighbors. This allows us to isolate the contribution of explicit graph modeling by comparing performance between Graphia and its sequential variant Graphia-seq.

We design two evaluation tasks targeting different levels of social dynamics:
\begin{itemize}
    \item \textbf{TDGG-task.} Evaluates the fidelity of local agent behaviors generated by LLMs.
    \item \textbf{IDGG-task.} Assesses system-level accuracy in predicting future social network evolution over time.
\end{itemize}

\noindent\textbf{Baselines for TDGG.} \ In the TDGG task, we examine both model scale and training paradigm. Evaluated models include:
\begin{itemize}
    \item Qwen3-8B, Qwen3-32B~\cite{yang2025qwen3}
    \item DeepSeek-R1-Distill-Qwen-32B~\cite{guo2025deepseek}
    \item Llama-3.1-70B-Instruct~\cite{dubey2024llama}
\end{itemize}
We additionally include a supervised fine-tuned version of Qwen3-8B (denoted Qwen3-SFT) to analyze the effect of direct behavioral cloning without reinforcement learning.

To study the impact of input data structure, we train Graphia-seq on the sequentialized dataset. The model follows the same architecture and training procedure as Graphia with sequential data. For reward design, we adopt established approaches from prior work on sequential data: destination selection is guided by a reward function adapted from LIKR~\cite{LIKR}, and edge generation is guided by a reward function adapted from Sotopia~\cite{zhousotopia}. 

\begin{table*}[t]
    \small
\centering
\caption{Metrics for the destination selection task. The best and second-best results are highlighted in \textbf{bold} and \underline{underline}, respectively.}
\label{tab:retrieval_metrics}
\begin{adjustbox}{width=.8\textwidth, totalheight=.8\textheight, keepaspectratio}
\begin{tabular}{ll|rrrrrrr}
\toprule
Dataset & Model & \multicolumn{2}{c}{Easy} & \multicolumn{2}{c}{Hard} & \multicolumn{2}{c}{All} \\
 &  & $\mathrm{H}$@100 & $\mathrm{R}$@100 & $\mathrm{H}$@100 & $\mathrm{R}$@100 & $\mathrm{H}$@100 & $\mathrm{R}$@100 \\
\midrule
\multirow{7}{*}{Propagate-En} & Qwen3-8b & 0.7450 & 0.4451 & 0.3275 & 0.3275 & 0.4550 & 0.3634 \\
 & Qwen3-8B-SFT  & \underline{0.7828} & 0.4601 & 0.3275 & 0.3275 & 0.4655 & 0.3677 \\
 & Qwen3-32B & 0.7606 & 0.4444 & \underline{0.3415} & \underline{0.3415} & 0.4648 & 0.3718 \\
 & DeepSeek-Q-32B & 0.7667 & \underline{0.4617} & 0.3125 & 0.3125 & 0.4515 & 0.3582 \\
 & Llama3.1-70B & 0.7513 & 0.4418 & \textbf{0.3437} & \textbf{0.3437} & \underline{0.4676} & \underline{0.3735} \\
 & Graphia-seq & 0.7449 & 0.4439 & 0.3297 & 0.3297 & 0.4547 & 0.3641 \\
 & Graphia & \textbf{0.7910} & \textbf{0.4763} & 0.3319 & 0.3319 & \textbf{0.4726} & \textbf{0.3761} \\
\midrule
\multirow{7}{*}{Weibo Tech} & Qwen3-8b & 0.5971 & 0
0.2602 & 0.2301 & 0.2301 & 0.4188 & 0.2455 \\
 & Qwen3-8B-SFT  & 0.5592 & 0.2422 & 0.2250 & 0.2250 & 0.3878 & 0.2334 \\
 & Qwen3-32B & \underline{0.6152} & 0.2641 & \underline{0.2346} & \underline{0.2346} & 0.4303 & 0.2498 \\
 & DeepSeek-Q-32B & \textbf{0.6246} & \textbf{0.2767} & 0.2235 & 0.2235 & \textbf{0.4372} & \underline{0.2518} \\
 & Llama3.1-70B & 0.6025 & 0.2607 & 0.2283 & 0.2283 & 0.4186 & 0.2448 \\
 & Graphia-seq & 0.6051 & 0.2629 & 0.2232 & 0.2232 & 0.4212 & 0.2438 \\
 & Graphia & \underline{0.6152} & \underline{0.2700} & \textbf{0.2364} & \textbf{0.2364} & \underline{0.4329} & \textbf{0.2538} \\
 \midrule
\multirow{7}{*}{Weibo Daily} & Qwen3-8b & 0.5800 & 0.3326 & 0.3030 & 0.3030 & 0.4011 & 0.3135 \\
 & Qwen3-8B-SFT  & 0.5453 & 0.3096 & 0.3060 & 0.3060 & 0.3850 & 0.3072 \\
 & Qwen3-32B & \underline{0.5827} & 0.3344 & \textbf{0.3159} & \textbf{0.3159} & \textbf{0.4123} & \textbf{0.3226} \\
 & DeepSeek-Q-32B & \textbf{0.5918} & \textbf{0.3401} & 0.2769 & 0.2769 & 0.3906 & 0.2997 \\
 & Llama3.1-70B & 0.5747 & 0.3259 & \underline{0.3127} & \underline{0.3127} & \underline{0.4044} & \underline{0.3173} \\
 & Graphia-seq & 0.5792 & 0.3325 & 0.3042 & 0.3042 & 0.4012 & 0.3142 \\
 & Graphia & 0.5800 & \underline{0.3379} & 0.3042 & 0.3042 & 0.4028 & 0.3162 \\
\bottomrule
\end{tabular}
\end{adjustbox}
\end{table*}

\noindent\textbf{Baselines for IDGG.} \ In the IDGG task, we benchmark against a range of representative social simulators spanning different modeling paradigms. This setup allows for a comparative analysis of approaches, from purely neural models to those incorporating LLM-based components:

\begin{itemize}
  \item \textbf{Deep Learning-Based Graph Generators.} These models are specifically designed to capture the evolution of temporal graph structures and serve as strong non-LLM baselines. (i) DGGen~\cite{DGGen}: A temporal GNN-based model for step-ahead graph prediction. (ii) TIGGER~\cite{TIGGER}: Uses probabilistic rules learned from historical interactions. (iii) DGGen variants: we further note that DGGen uses TGN~\cite{TGN} as its backbone to learn temporal graph embeddings. Following the DGGen framework, we replace the backbone with more recent temporal graph learning models~\cite{DTGB}, yielding DGGen variants with GraphMixer~\cite{graphmixer}, CAWN~\cite{CAWN}, and Dygformer~\cite{Dygformer} backbones.
  \item \textbf{Hybrid LLM-based Social Simulators.} We implement a baseline based on SAGraph~\cite{sagraph_bench}, which predicts daily activity levels by fitting Gaussian distributions to historical node activity. The predicted number of active edges determines how many source nodes are sampled per day. Given active source nodes, we use two LLMs to generate interaction: Qwen3-SFT and Graphia-seq.
  \item \textbf{Pure LLM-Based Simulators:} We also compare with GAG-General~\cite{peng2025gdgb}, a recent LLM-based system designed for general-purpose graph generation. It uses prompt engineering and ReAct to simulate agent decisions and network growth. 
  We set the seed graph length to 10000 edges, (closet to the prediction time), We implement GAG-General based on Llama3-8B~\footnote{https://huggingface.co/meta-llama/Meta-Llama-3-8B} and Qwen3-8B~\footnote{https://huggingface.co/Qwen/Qwen3-8B} backbone.
\end{itemize}

We train all baseline models on the training split and generate future graphs matching the temporal extent of the test split. Since GAG-General only provides graph generation pipelines for Weibo Tech and Weibo Daily, we do not report its performance on the Propagate-En dataset.

\begin{table*}[htbp]
  \centering
  % \small
  \caption{LLM-as-a-judge for the edge generation task, we adopt Qwen2-72B as the evaluation LLM. The best and second-best results are highlighted in \textbf{bold} and \underline{underline}, respectively.}
  \label{tab:edge_llm_eval_all}
  \begin{adjustbox}{width=.8\textwidth, totalheight=.8\textheight, keepaspectratio}
  \begin{tabular}{ll|rrrrrrr}
    \toprule
    Dataset & Model & GF & CF & PD & DA & IQ & CR & Average \\
    \midrule
    % \multirow{4}{*}{Propagate-En} 
    %  & Qwen3-8B & 1.9106 & 1.8655 & 1.8322 & 1.8541 & 1.8708 & 1.8708 & 1.8674 \\
    %  & Qwen3-8B-SFT  & \underline{2.7647} & \underline{2.7647} & \underline{3.8235} & \underline{2.7647} & \underline{2.7647} & \textbf{2.8824} & \underline{2.9608} \\
      
    %  & Graphia-seq & 1.6995 & 1.6978 & 2.035 & 1.6916 & 1.6982 & 1.7000 & 1.7537 \\
    %  & Graphia & \textbf{2.7877} & \textbf{2.7877} & \textbf{3.9147} & \textbf{2.7817} & \textbf{2.7857} & \underline{2.7996} & \textbf{2.9762} \\
    % \midrule
    % \multirow{4}{*}{Weibo Daily} 
    %  & Qwen3-8B & 2.4834 & 2.4859 & 2.482 & 2.4541 & 2.48 & 2.4453 & 2.4718 \\
    %  & Qwen3-8B-SFT & \underline{2.5405} & \underline{2.5427} & \underline{2.5157} & \underline{2.5226} & \underline{2.5382} & \underline{2.5114} & \underline{2.5285} \\
     
    %  & Graphia-seq & 2.4694 & 2.4715 & 2.4711 & 2.4406 & 2.4656 & 2.4326 & 2.4585 \\
    %  & Graphia & \textbf{4.2161} & \textbf{4.2177} & \textbf{4.1638} & \textbf{4.1799} & \textbf{4.212} & \textbf{4.1711} & \textbf{4.1934} \\
    % \midrule
    % \multirow{4}{*}{Weibo Tech} 
    %  & Qwen3-8B & 2.4308 & 2.432 & 2.4192 & 2.3941 & 2.4267 & 2.3806 & 2.4139 \\
    %  & Qwen3-8B-SFT  & \underline{2.7564} & \underline{2.7597} & \underline{2.7401} & \underline{2.7379} & \underline{2.7538} & \underline{2.7223} & \underline{2.745} \\
     
    %  & Graphia-seq & 2.4365 & 2.4388 & 2.423 & 2.3978 & 2.4314 & 2.3833 & 2.4185 \\
    %  & Graphia & \textbf{3.3692} & \textbf{3.3705} & \textbf{3.3449} & \textbf{3.3478} & \textbf{3.3666} & \textbf{3.3343} & \textbf{3.3555} \\

    \multirow{6}{*}{Propagate-En}  
    & Qwen3 & 1.9106 & 1.8655 & 1.8322 & 1.8541 & 1.8708 & 1.8708 & 1.8674 \\
  & Qwen-3-SFT & 2.7647 & 2.7647 & 3.8235 & \underline{2.7647} & 2.7647 & \textbf{2.8824} & \underline{2.9608} \\
  & Qwen3-8B+TGN & 2.6167 & 2.6180 & 3.5817 & 2.5948 & 2.6180 & 2.6211 & 2.7751 \\
  & Qwen3-8B+DyGFormer & 2.6307 & 2.6303 & 3.5738 & 2.6124 & 2.6290 & 2.6321 & 2.7847 \\
  & Qwen3-8B+GraphMixer & \underline{2.7665} & \underline{2.7678} & \underline{3.8445} & 2.7464 & \underline{2.7678} & 2.7700 & 2.9439 \\
  & Graphia & \textbf{2.7877} & \textbf{2.7877} & \textbf{3.9147} & \textbf{2.7817} & \textbf{2.7857} & \underline{2.7996} & \textbf{2.9762} \\

  \midrule
    
  \multirow{6}{*}{Weibo Tech}  
  & Qwen3 & 2.4308 & 2.4320 & 2.4192 & 2.3941 & 2.4267 & 2.3806 & 2.4139 \\
  & Qwen3-8B-SFT & \underline{2.7564} & \underline{2.7597} & \underline{2.7401} & \underline{2.7379} & \underline{2.7538} & \underline{2.7223} & \underline{2.7450} \\
  & Qwen3-8B+TGN & 2.3660 & 2.3669 & 2.3559 & 2.3576 & 2.3631 & 2.3504 & 2.3600 \\
  & Qwen3-8B+DyGFormer & 2.3697 & 2.3707 & 2.3626 & 2.3604 & 2.3684 & 2.3550 & 2.3645 \\
  & Qwen3-8B+GraphMixer & 2.3110 & 2.3116 & 2.3062 & 2.3029 & 2.3094 & 2.2945 & 2.3059 \\
  & Graphia & \textbf{3.3692} & \textbf{3.3705} & \textbf{3.3449} & \textbf{3.3478} & \textbf{3.3666} & \textbf{3.3343} & \textbf{3.3555} \\

  \midrule

   \multirow{6}{*}{Weibo Daily}  
   & Qwen3 & 2.4834 & 2.4859 & 2.4820 & 2.4541 & 2.4800 & 2.4453 & 2.4718 \\
  & Qwen3-8B-SFT & \underline{2.5405} & \underline{2.5427} & \underline{2.5157} & \underline{2.5226} & \underline{2.5382} & \underline{2.5114} & \underline{2.5285} \\
  & Qwen3-8B+TGN & 2.4066 & 2.4083 & 2.3925 & 2.4012 & 2.4058 & 2.3983 & 2.4021 \\
  & Qwen3-8B+DyGFormer & 2.4560 & 2.4571 & 2.4389 & 2.4489 & 2.4543 & 2.4466 & 2.4503 \\
  & Qwen3-8B+GraphMixer & 2.4192 & 2.4207 & 2.4044 & 2.4123 & 2.4178 & 2.4107 & 2.4142 \\
  & Graphia & \textbf{4.2161} & \textbf{4.2177} & \textbf{4.1638} & \textbf{4.1799} & \textbf{4.2120} & \textbf{4.1711} & \textbf{4.1934} \\

    \bottomrule
  \end{tabular}
\end{adjustbox}
\end{table*}

% This hybrid social simulator evaluates the benefit of integrating LLM-driven interaction generation within traditional sociodynamic frameworks.

\section{Supplementary Experiments}

\subsection{TDGG Experiments}

\noindent\textbf{Destination Selection.} \ As shown in Table~\ref{tab:retrieval_metrics}, we report Hit@100 ($\mathrm{H}$@100) and Recall@100 ($\mathrm{R}$@100) for destination selection on three datasets: Propagate-En, Weibo Daily, and Weibo Tech.

On Propagate-En, Graphia achieves the highest performance on the full test set with $\mathrm{H}$@100 = 0.7910 and $\mathrm{R}$@100 = 0.4763, outperforming both the base Qwen3-8B model and its SFT variant. It also surpasses larger models such as Qwen3-32B and Llama3.1-70B, suggesting that the RL-based training contributes to improved prediction accuracy. Gains are modest on the hard subset but more evident in the easy and overall settings.
On Weibo Daily, Qwen3-32B ($\mathrm{H}$@100 = 0.5827) and DeepSeek-Q-32B ($\mathrm{H}$@100 = 0.5918) achieve the best Hit@100 scores. Graphia performs comparably in $\mathrm{H}$@100 (0.5800) and attains the highest $\mathrm{R}$@100 (0.3379), indicating slightly better coverage of true destinations despite similar ranking performance.
On Weibo Tech, Graphia matches the top $\mathrm{H}$@100 score (0.6152, shared with Qwen3-32B), achieves the highest $\mathrm{R}$@100 (0.2700), and performs best on the hard subset ($\mathrm{H}$@100 = 0.2364), suggesting effectiveness in identifying interaction partners.

The Graphia-seq variant, which is trained using reinforcement learning on sequential data without incorporating structural feedback, performs similarly to the SFT baseline. The superior performance of Graphia underscores that incorporating graph-structured data can effectively boost LLMs' ability to select appropriate interaction partners.

\noindent\textbf{Edge Generation.} \ 
Structural prediction assesses whether the generated graph structure aligns with the reference graph, but textual content is the core focus in our evaluation. To strengthen the evaluation of Graphia's ability to generate interaction content, we introduce a new baseline: DGNN+LLM. Specifically, we leverage the state-of-the-art dynamic graph neural networks (DGNN) from DTGB~\cite{DTGB}: TGN~\cite{TGN}, DyGFormer~\cite{Dygformer}, GraphMixer~\cite{graphmixer}. The DGNNs are used to predict edge types, and LLMs are used to generate the edge content for predicted edge type.

Table~\ref{tab:edge_llm_eval_all} presents LLM-as-a-judge scores across six dimensions: Goal Fulfillment (GF) from SOTOPIA~\cite{zhousotopia}, Contextual Fidelity (CF), Personality Depth (PD), Dynamic Adaptability (DA), Immersive Quality (IQ), and Content Richness (CR) from GDGB~\cite{peng2025gdgb} for edge message generation on Propagate-En, Weibo Daily, and Weibo Tech.

First, the results suggest that edge generation is a relatively accessible task compared to destination selection. Even the base Qwen3-8B model achieves moderate performance, with average scores ranging from 2.41 to 2.47 on the Weibo datasets and 1.87 on Propagate-En. Moreover, DGNN provides some auxiliary benefit to the LLM: on the Propagate-EN dataset, DGNN+LLM achieves text quality comparable to the SFT-finetuned model, but fails to improve performance on Weibo-Daily and Weibo-Tech. In comparison, Graphia achieves the best performance on all three datasets.These values are substantially higher than typical baseline performance in node retrieval tasks, indicating that generating plausible interaction text benefits heavily from pre-trained language priors.

Second, despite the low barrier to entry for LLM-based agents, structured training plays a critical role in performance. The Graphia-seq variant, which uses only sequential interaction data without explicit topological modeling, performs worse than the SFT baseline in most cases. In contrast, Graphia, which incorporates both graph-structural context and structured reward modeling during reinforcement learning, achieves significant improvements, with an average score improvement of 1.66 over SFT on Weibo Daily and 0.61 on Weibo Tech. These gains are consistent across all evaluation dimensions, demonstrating that grounding agent behavior in structural dynamics leads to better contextual understanding and more socially coherent interactions.
\begin{figure}[htbp]
  \subfloat{\includegraphics[width=\linewidth]{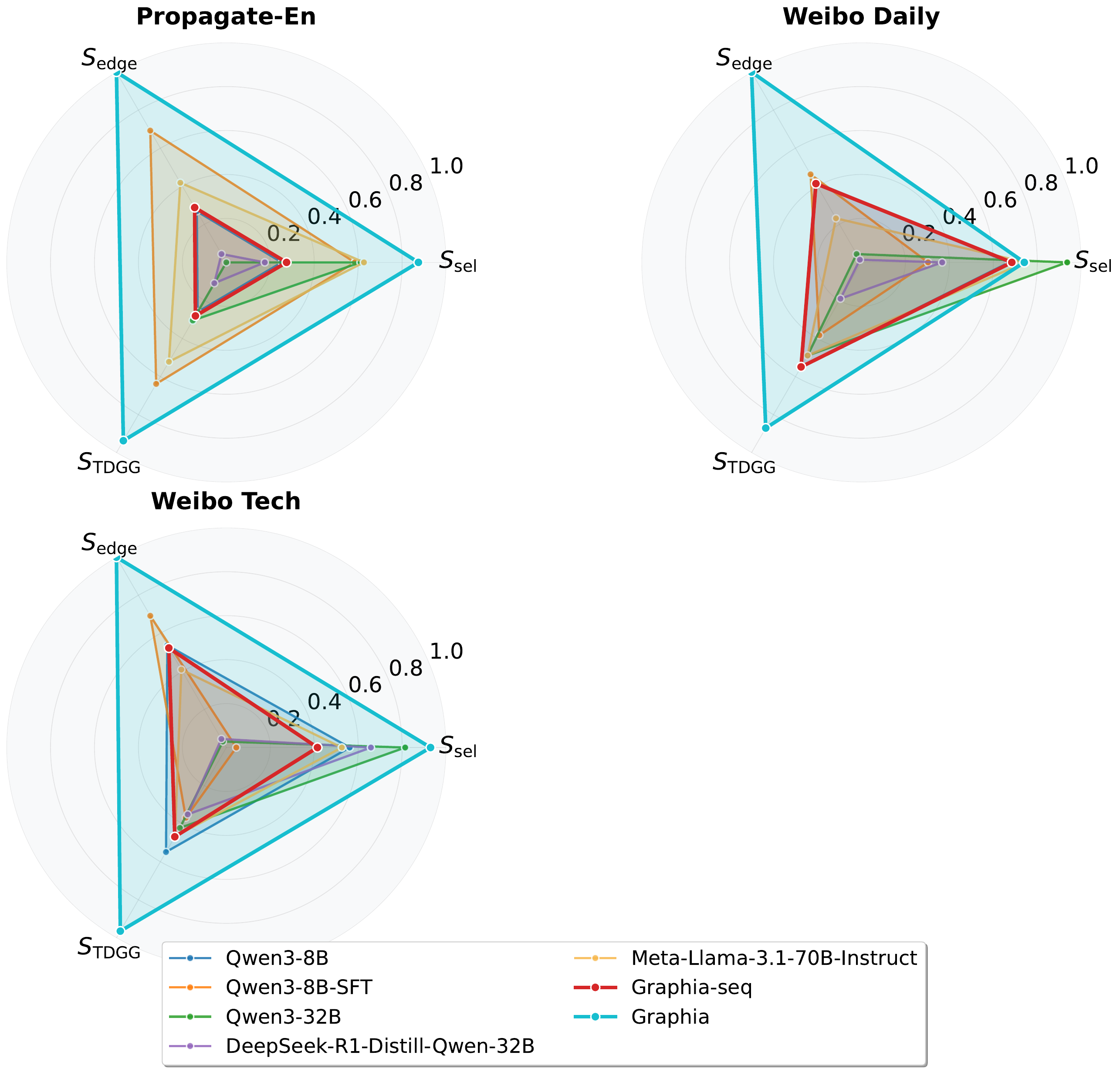
  }}
  % \vspace{1em}
\caption{The social fidelity score for the TDGG task on three social network datasets.}
\label{fig:tdgg_details}
\end{figure}

\noindent\textbf{Overall Comparison.} \ As shown in Figure~\ref{fig:tdgg_details}, we report the destination selection score ($S_\text{sel}$), edge generation score ($S_\text{edge}$), and overall TDGG performance ($S_\text{TDGG}$) across multiple models on three datasets.
Compared to its base model Qwen3-8B, Graphia achieves substantial improvements in edge generation through reinforcement learning on social graph data, while yielding more modest gains in destination selection. This suggests that generating semantically coherent edges is more readily optimized via reward-guided training than accurately predicting high-level node destinations.

When comparing all models, the ranking in $S_\text{sel}$ is approximately: Graphia, Qwen3-32B, Llama-3.1-70B-Instruct, DeepSeek-R1-Distill-Qwen-32B, Qwen3-8B, and Qwen3-8B-SFT. Larger-parameter models generally outperform smaller ones, indicating that destination selection benefits from increased model capacity and world knowledge.
In contrast, the ranking for $S_\text{edge}$ follows a different pattern: Graphia, Qwen3-8B-SFT, Qwen3-8B, Graphia-seq, Llama-3.1-70B-Instruct, DeepSeek-R1-Distill-Qwen-32B, and Qwen3-32B. Notably, fine-tuned smaller models outperform even the largest LLMs, and Graphia significantly surpasses all baselines. This highlights the effectiveness of RL-based alignment with structural feedback in improving edge content quality beyond what scale or supervised fine-tuning alone can achieve.

These results indicate that while edge generation can be effectively enhanced through targeted reinforcement learning, destination selection remains a more challenging task that demands deeper reasoning over long-term human behavior. This finding aligns with recent observations in PersonaEval~\cite{zhou2025personaeval}.

\subsection{Ablation Experiment on Filter}
As shown in Table~\ref{tab:filter_ablation}, applying a post-hoc filter to Graphia's generated edges leads to a measurable improvement in destination selection performance. The gain primarily comes from enforcing that nodes sharing many common neighbors are more likely to be connected, which is a structural property commonly observed in real-world graphs.

\subsection{Ablation Experiment on Reward}
To evaluate the effectiveness of our reward design, we conduct ablation studies on the components of the reward function. 
For Graphia(w/o cat), we remove the category prediction reward $r_{\text{cat}}$ during training and use only the text generation reward $r_{\text{text}}$. 
For Graphia(w/o GNN), we retain the dual-domain, domain-interleaved sampling strategy with a fixed 1:1 ratio (category:message), but remove the GNN-based structural reward component. Specifically, we modify $r_{\text{cat}}$ to:
\[
r_{\text{cat}} = \mathbb{I}(\hat{y}_{u\to v} = y^*) + r_{\text{format}},
\]
where the indicator term rewards correct category prediction and $r_{\text{format}}$ ensures output formatting correctness, without incorporating graph-level consistency signals from the GNN. We train the variants using the same number of steps as Graphia-E (100 steps).

As shown in Table~\ref{tab:ablation_reward_edge}, for the Weibo-Tech dataset, compared to the base models Qwen3-8B and Qwen3-8B-SFT, we observe that supervised fine-tuning alone improves category prediction accuracy by approximately 10\%. However, Graphia(w/o cat) achieves only a marginal gain of 0.5\% over Qwen3-8B-SFT; Graphia(w/o GNN) achieves a 4.65\% increase in accuracy over Qwen3-8B-SFT. 
As shown in Table~\ref{tab:ablation_reward_edge_weibo_daily}, for the Weibo-Daily dataset, a similar trend is observed. Qwen3-8B-SFT improves accuracy by nearly 9\% over Qwen3-8B. Graphia(w/o cat) yields a modest 1.85\% improvement over Qwen3-8B-SFT, Graphia(w/o GNN) attains a 24.63\% increase in accuracy over Qwen3-8B-SFT; Graphia still achieves the best performance with 24.98\% over Qwen3-8B-SFT.
This demonstrate that optimizing $r_{\text{cat}}$ during reinforcement learning provides measurable benefits. Nevertheless, Graphia which incorporates GNN-as-reward, delivers a substantial 8.96\% improvement in category prediction accuracy, highlighting the critical role of graph-structural feedback in aligning agent behavior with ground-truth interaction patterns.

\begin{table}[htbp]
\centering
\caption{ Ablation study on the effect of applying the filtering mechanism during evaluation for the destination selection task.}

\label{tab:filter_ablation}
\begin{adjustbox}{width=.5\textwidth, totalheight=.5\textheight, keepaspectratio}
\begin{tabular}{lrrrr}
\toprule
 & H@100-ALL &  & R@100-ALL &  \\
& w/ filter & wo. filter & w/ filter & wo. filter \\
\midrule
% \multirow{2}{*}{Propagate-En} & Qwen3 & 0.46 & 0.13 & 0.36 & 0.08 \\
% & Graphia & 0.47 & 0.25 & 0.38 & 0.18 \\
% \multirow{2}{*}{Weibo Tech} & Qwen3 & 0.42 & 0.08 & 0.25 & 0.02 \\
% & Graphia && 0.43 & 0.08 & 0.25 & 0.01 \\
% \multirow{2}{*}{Weibo Daily}& Qwen3 & 0.40 & 0.08 & 0.31 & 0.05 \\
% & Graphia & 0.40 & 0.04 & 0.32 & 0.01 \\
\multicolumn{5}{c}{\textbf{Propagate-En}} \\
Qwen3-8B & 0.455 & 0.127 & 0.363 & 0.077 \\
Graphia & 0.473 & 0.253 & 0.376 & 0.178 \\
\midrule
 \multicolumn{5}{c}{\textbf{Weibo Tech}} \\
Qwen3-8B & 0.419 & 0.079 & 0.246 & 0.021 \\
Graphia & 0.433 & 0.079 & 0.254 & 0.010 \\
\midrule
 \multicolumn{5}{c}{\textbf{Weibo Daily}} \\
Qwen3-8B & 0.401 & 0.078 & 0.314 & 0.045 \\
Graphia & 0.403 & 0.035 & 0.316 & 0.012 \\
\bottomrule
\end{tabular}
\end{adjustbox}
\end{table}

\begin{table}
\caption{Performance comparison of Graphia variants on the Weibo Tech dataset for the edge generation task. The best and second-best results are highlighted in \textbf{bold} and \underline{underline}, respectively.}
\label{tab:ablation_reward_edge}
\small
\begin{adjustbox}{width=.5\textwidth, totalheight=.5\textheight, keepaspectratio}
\begin{tabular}{l|rrr}
\toprule
Model & $\mathrm{ACC}$ $\uparrow$ & ROUGE-L $\uparrow$ & BERTScore $\uparrow$ \\
\midrule
Qwen3-8B & 0.6326 & 0.6014 & 0.1757 \\
Qwen3-8B-SFT & 0.7325 & 0.5985 & 0.2128 \\
Graphia(w/o cat) & 0.7362 & 0.6012 & 0.2111 \\
Graphia(w/o GNN) & \underline{0.7790} & \underline{0.6018} & \underline{0.2737} \\
Graphia & \textbf{0.8221} & \textbf{0.6040} & \textbf{0.2963} \\

\bottomrule
\end{tabular}
\end{adjustbox}
\end{table}

\begin{table}
\caption{Performance comparison of Graphia variants on the Weibo Daily dataset for the edge generation task. The best and second-best results are highlighted in \textbf{bold} and \underline{underline}, respectively.}
\label{tab:ablation_reward_edge_weibo_daily}
\small
\begin{adjustbox}{width=.5\textwidth, totalheight=.5\textheight, keepaspectratio}
\begin{tabular}{l|rrr}
\toprule
Model & $\mathrm{ACC}$ $\uparrow$ & ROUGE-L $\uparrow$ & BERTScore $\uparrow$ \\
\midrule
Qwen3-8B & 0.5456 & 0.5661 &	0.1243 \\
Qwen3-8B-SFT & 0.6338 & 0.5755 & 0.0735 \\
Graphia(w/o cat) & 0.6523 & 0.5723 & 0.0260 \\
Graphia(w/o GNN) & \underline{0.8801} & \underline{0.5901} & \underline{0.1594} \\
Graphia & \textbf{0.8836} & \textbf{0.6088} & \textbf{0.2652} \\

\bottomrule
\end{tabular}
\end{adjustbox}
\end{table}

\subsection{Ablation on Evaluation LLMs}
\label{appendix:llm_judge_variants}
% TBD: add explanation: llama31经常爆0，llama系列格式错误太多，导致分数整体低于qwen系列.

To mitigate potential bias arising from using a reward model (Qwen3-8B) that may favor outputs in its own linguistic style—thereby introducing unfairness in evaluation, we conduct an ablation study on the impact of different LLM-as-a-judge models on the final assessment. Specifically, we select four distinct large language models as judges: Llama-3.1-70B, Llama-3.3-70B, Qwen3-32B, and Qwen2-72B. These models independently evaluate the edge generation task. The results are summarized in Table~\ref{tab:eval_llm_comparison}.

As shown in the table, while the absolute scores vary across different judges, the overall performance trend remains consistent: \textsc{Graphia} consistently outperforms both Qwen3 and Qwen3-SFT across the majority of settings. This demonstrates that \textsc{Graphia}'s advantage in edge generation is robust and not dependent on a particular evaluation model. Notably, we observe that Llama-3.1-70B frequently produces malformed outputs during evaluation, leading to lower and less reliable scores. This highlights the importance of selecting stable and well-behaved models when deploying LLM-as-a-judge protocols.

\begin{table*}[htbp]
\centering
\small
\caption{ Ablation study on different LLMs-as-a-judge. We report the scores on six dimensions: Goal Fulfillment (GF), Contextual Fidelity (CF), Personality Depth (PD), Dynamic Adaptability (DA), Immersive Quality (IQ), and Content Richness (CR). The best and second-best results are highlighted in \textbf{bold} and \underline{underline}, respectively.}
\label{tab:eval_llm_comparison}
\begin{tabular}{l|l|rrrrrrr}
\toprule
LLM-as-a-judge & Model & GF & CF & PD & DA & IQ & CR & Average \\
\midrule
\multicolumn{9}{c}{\textbf{Propagate-En}} \\
\midrule
\multirow{3}{*}{Llama31-70B} & Qwen3 & 1.6216 & 1.5747 & 1.4630 & 1.3049 & 1.5164 & 1.4520 & 1.4888 \\
 & Qwen3-SFT & \underline{2.0280} & \textbf{1.9711} & \textbf{1.7823} & \textbf{1.5852} & \textbf{1.8747} & \textbf{1.8046} & \textbf{1.8410} \\
 & Graphia & \textbf{2.0368} & \underline{1.9483} & \underline{1.7407} & \underline{1.5331} & \underline{1.8625} & \underline{1.7613} & \underline{1.8138} \\
% \cline{3-9} 
 \multirow{3}{*}{Llama33-70B} & Qwen3 & 1.5103 & 1.4700 & 1.3631 & \textbf{1.1993} & \textbf{1.1756} & \textbf{1.0166} & \underline{1.2892} \\
 & Qwen3-SFT & \underline{1.6364} & \underline{1.5629} & \underline{1.4503} & 1.1253 & 0.9912 & 0.7718 & 1.2563 \\
 & Graphia & \textbf{1.6575} & \textbf{1.6001} & \textbf{1.4792} & \underline{1.1848} & \underline{1.0320} & \underline{0.7819} & \textbf{1.2892} \\
\multirow{3}{*}{Qwen3-32B} & Qwen3 & 1.8647 & 1.8287 & 1.5090 & 1.5545 & 1.6404 & 1.5957 & 1.6655 \\
 & Qwen3-SFT & \textbf{2.2698} & \underline{2.3031} & \underline{1.7367} & \textbf{1.8760} & \underline{1.9242} & \textbf{1.9474} & \underline{2.0096} \\
 & Graphia & \underline{2.2523} & \textbf{2.3149} & \textbf{1.7429} & \underline{1.8585} & \textbf{1.9439} & \underline{1.9461} & \textbf{2.0098} \\
\multirow{3}{*}{Qwen2-72B} & Qwen3 & 1.9106 & 1.8655 & 1.8322 & 1.8541 & 1.8708 & 1.8708 & 1.8674 \\
 & Qwen3-SFT & \underline{2.7647} & \underline{2.7647} & \underline{3.8235} & \underline{2.7647} & \underline{2.7647} & \textbf{2.8824} & \underline{2.9608} \\
 & Graphia & \textbf{2.7877} & \textbf{2.7877} & \textbf{3.9147} & \textbf{2.7817} & \textbf{2.7857} & \underline{2.7996} & \textbf{2.9762} \\
\midrule
\multicolumn{9}{c}{\textbf{Weibo Tech}} \\
\midrule
\multirow{3}{*}{Llama31-70B} & Qwen3 & 1.7852 & 1.6641 & 1.5688 & 1.3179 & 1.6105 & 1.4545 & 1.5668 \\
 & Qwen3-SFT & \underline{1.9383} & \underline{1.8265} & \underline{1.7613} & \underline{1.5155} & \underline{1.7844} & \underline{1.6821} & \underline{1.7514} \\
 & Graphia & \textbf{2.1057} & \textbf{1.9651} & \textbf{1.9069} & \textbf{1.6036} & \textbf{1.9119} & \textbf{1.8324} & \textbf{1.8876} \\
\multirow{3}{*}{Llama33-70B} & Qwen3 & \underline{2.2075} & \underline{2.0964} & 1.8625 & 1.6595 & \underline{1.9860} & 1.7091 & 1.9202 \\
 & Qwen3-SFT & 2.1691 & 2.0832 & \underline{1.9698} & \underline{1.8241} & 1.9356 & \underline{1.7682} & \underline{1.9583} \\
 & Graphia & \textbf{2.4125} & \textbf{2.3022} & \textbf{2.1855} & \textbf{2.0081} & \textbf{2.1365} & \textbf{1.9416} & \textbf{2.1644} \\
\multirow{3}{*}{Qwen3-32B} & Qwen3 & 2.3356 & 2.3640 & 1.9314 & 2.0303 & 2.1692 & 1.8684 & 2.1165 \\
 & Qwen3-SFT & \underline{2.7140} & \underline{2.7404} & \underline{2.3013} & \underline{2.4148} & \underline{2.5553} & \underline{2.2794} & \underline{2.5009} \\
 & Graphia & \textbf{3.2165} & \textbf{3.2577} & \textbf{2.7452} & \textbf{2.8738} & \textbf{3.0502} & \textbf{2.7412} & \textbf{2.9808} \\
\multirow{3}{*}{Qwen2-72B} & Qwen3 & 2.4308 & 2.4320 & 2.4192 & 2.3941 & 2.4267 & 2.3806 & 2.4139 \\
 & Qwen3-SFT & \underline{2.7564} & \underline{2.7597} & \underline{2.7401} & \underline{2.7379} & \underline{2.7538} & \underline{2.7223} & \underline{2.7450} \\
 & Graphia & \textbf{3.3692} & \textbf{3.3705} & \textbf{3.3449} & \textbf{3.3478} & \textbf{3.3666} & \textbf{3.3343} & \textbf{3.3555} \\
\midrule
\multicolumn{9}{c}{\textbf{Weibo Daily}} \\
\midrule
\multirow{3}{*}{Llama31-70B} & Qwen3 & 1.9201 & 1.7534 & 1.6956 & 1.3937 & 1.6998 & 1.5490 & 1.6686 \\
 & Qwen3-SFT & \underline{2.1431} & \underline{1.9504} & \underline{1.9060} & \underline{1.5975} & \underline{1.9067} & \underline{1.7533} & \underline{1.8762} \\
 & Graphia & \textbf{3.1950} & \textbf{2.8984} & \textbf{2.7752} & \textbf{2.2418} & \textbf{2.8354} & \textbf{2.6460} & \textbf{2.7653} \\
\multirow{3}{*}{Llama33-70B} & Qwen3 & \underline{2.7762} & \underline{2.6386} & \underline{2.4059} & \underline{2.1984} & \underline{2.4679} & \underline{2.2207} & \underline{2.4513} \\
 & Qwen3-SFT & 2.3057 & 2.2256 & 2.1454 & 1.9926 & 2.0469 & 1.9035 & 2.1033 \\
 & Graphia & \textbf{3.6851} & \textbf{3.5156} & \textbf{3.0989} & \textbf{2.8741} & \textbf{3.2346} & \textbf{2.8684} & \textbf{3.2128} \\
\multirow{3}{*}{Qwen3-32B} & Qwen3 & 2.3332 & 2.3215 & 1.9612 & 2.0209 & 2.1872 & 1.9146 & 2.1231 \\
 & Qwen3-SFT & \underline{2.3845} & \underline{2.3751} & \underline{2.0603} & \underline{2.1335} & \underline{2.2671} & \underline{2.0380} & \underline{2.2097} \\
 & Graphia & \textbf{3.9917} & \textbf{4.0042} & \textbf{3.2503} & \textbf{3.4254} & \textbf{3.8050} & \textbf{3.2621} & \textbf{3.6231} \\
\multirow{3}{*}{Qwen2-72B} & Qwen3 & 2.4834 & 2.4859 & 2.4820 & 2.4541 & 2.4800 & 2.4453 & 2.4718 \\
 & Qwen3-SFT & \underline{2.5405} & \underline{2.5427} & \underline{2.5157} & \underline{2.5226} & \underline{2.5382} & \underline{2.5114} & \underline{2.5285} \\
 & Graphia & \textbf{4.2161} & \textbf{4.2177} & \textbf{4.1638} & \textbf{4.1799} & \textbf{4.2120} & \textbf{4.1711} & \textbf{4.1934} \\
\bottomrule
\end{tabular}
\end{table*}

\subsection{IDGG Experiments}
\begin{table*}[htbp]
\centering
\caption{Node degree prediction metrics for different datasets. The best and second-best results are highlighted in \textbf{bold} and \underline{underline}, respectively.}
\label{tab:activity_prediction}
\begin{adjustbox}{width=.77\textwidth, totalheight=.77\textheight, keepaspectratio}
\begin{tabular}{ll|rrr}
\toprule
Dataset & Model & Wasserstein Distance $\downarrow$ & KL-Divergence $\downarrow$ & $\mathrm{MMD.OD}$ $\downarrow$ \\
\midrule
\multirow{3}{*}{Propagate-En} 
  & DGGen & \underline{0.0152} & \textbf{1.0525} & 0.0016 \\
 & Tigger & 0.0225 & \underline{1.2222} & \textbf{0.0006} \\
 & Graphia & \textbf{0.0109} & 3.2884 & \underline{0.0013} \\
\midrule
\multirow{4}{*}{Weibo Tech} 
& GAG-General & 0.6696 & 19.3922 & \underline{0.0008} \\
 & DGGen & \underline{0.0803} & \underline{6.6602} & 0.0040 \\
 & Tigger & 0.0840 & 10.5437 & 0.0032 \\
 & Graphia & \textbf{0.0338} & \textbf{2.7597} & \textbf{0.0003} \\
 \midrule
\multirow{4}{*}{Weibo Daily} 
  & GAG-General & 0.4921 & 11.3634 & 0.0012 \\
 & DGGen & 0.2023 & 20.4695 & 0.0036 \\
 & Tigger & \underline{0.0909} & \underline{5.4737} & \underline{0.0006} \\
 & Graphia & \textbf{0.0134} & \textbf{1.0696} & \textbf{0.0003} \\
\bottomrule
\end{tabular}
\end{adjustbox}
\end{table*}

\noindent\textbf{Activity Prediction.} \ As shown in Table~\ref{tab:activity_prediction}, we evaluate the accuracy of source node out-degree prediction by binning degree values and computing distributional distances between the generated and reference graphs. The metrics of Wasserstein distance, KL-divergence, and MMD are all lower-is-better ($\downarrow$), indicating how closely the predicted degree distribution matches the ground truth.

\begin{figure}[htbp]
  \subfloat{\includegraphics[width=\linewidth]{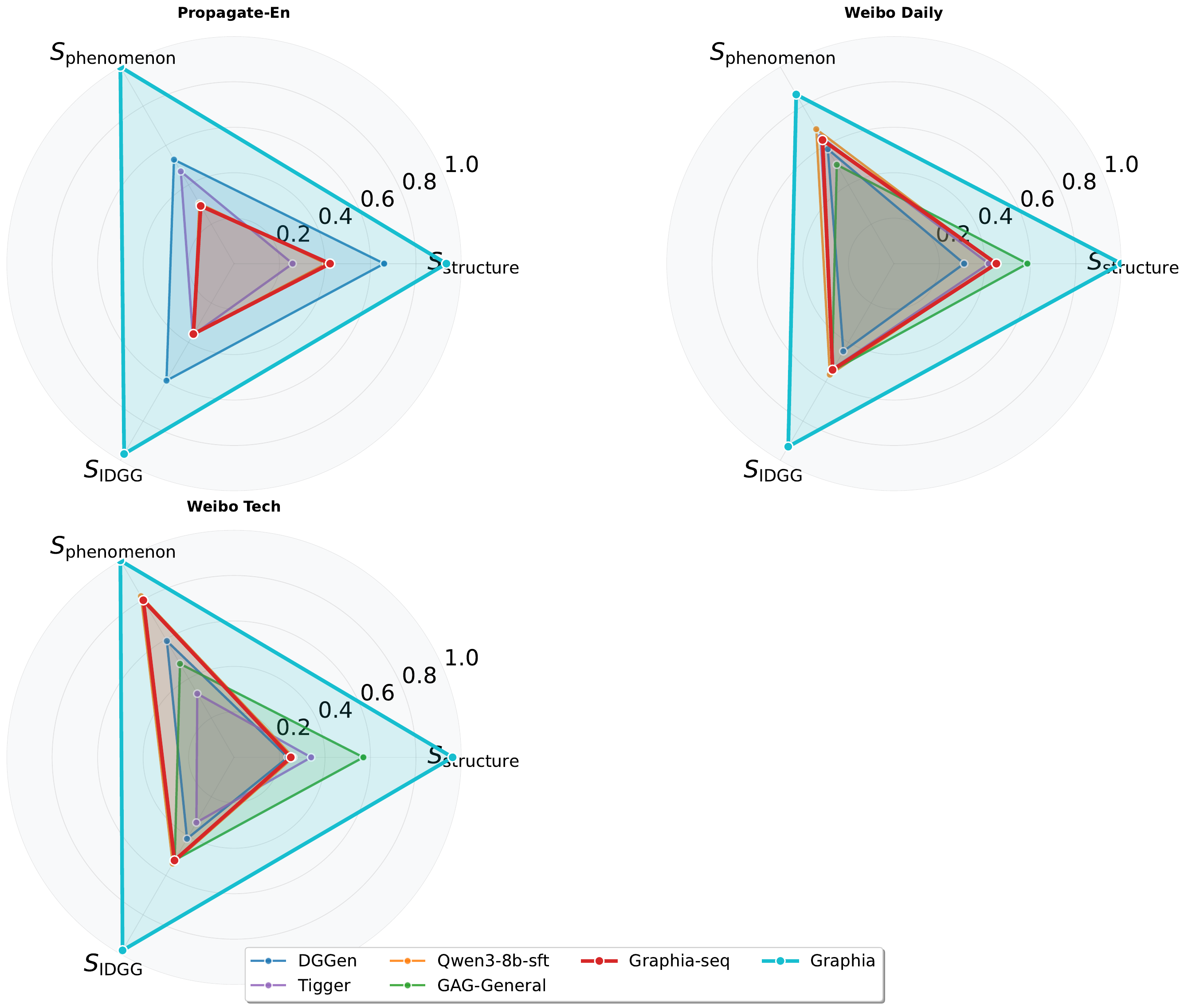
  }}
  % \vspace{1em}
\caption{The social fidelity score for the IDGG task on three social network datasets.}
\label{fig:idgg_details}
\end{figure}

The results show that Graphia's activity predictor significantly outperforms DGGen and GAG-General, which rely on random sampling strategies for node activation. It also surpasses TIGGER, a method based on temporal point processes for modeling event timing. This performance advantage arises from Graphia's structure-aware design: the activity predictor is built upon an Informer architecture~\cite{zhou2021informer} that explicitly integrates historical interaction patterns of evolving node degrees. By jointly modeling temporal dependencies and topological signals, Graphia captures more realistic user engagement dynamics, leading to more accurate activation patterns in social simulations.

\noindent\textbf{Overall Comparison.} \  We report the Macro Structure Fidelity score ($S_\text{structure}$), Macro Phenomenon Consistency score ($S_\text{phenomenon}$), and overall IDGG performance ($S_\text{IDGG}$) across multiple models on three datasets in Figure~\ref{fig:idgg_details}. We select GAG-General with Llama3 backbone for demonstration, as it outperforms the Qwen3 variant.
Overall, Graphia significantly outperforms all baselines in both structural fidelity and phenomenological consistency. By integrating reinforcement learning with structural feedback, Graphia achieves the top rank in $S_\text{IDGG}$ on all three datasets. This demonstrates its ability to generate dynamic graphs that simultaneously align with macroscopic topological properties and capture emergent social phenomena.

Notably, DGGen is a purely structure-driven dynamic graph generator that does not utilize textual content. It achieves the best performance among non-LLM approaches, highlighting the effectiveness of dynamic graph neural network architectures in preserving topological dynamics. In contrast, among all baselines, GAG-General and Qwen3-8B-SFT emerge as the strongest performers, surpassing traditional deep-learning-based models such as DGGen and TIGGER. This underscores the advantage of LLM-based approaches in capturing high-level interaction patterns and their potential for generating realistic social graphs.

\begin{table*}[htbp]
  \centering
  \caption{Ablation study on the training components of Graphia. We compare the effects of reinforcement learning (RL) and activity predictor (AP) training across three datasets. The best and second-best results are highlighted in \textbf{bold} and \underline{underline}, respectively.}
  \begin{adjustbox}{width=.9\textwidth, totalheight=\textheight, keepaspectratio}
  \begin{tabular}{l|rrrr|rrr}
    
    % \toprule
    % \textbf{model} & \textbf{degree mmd} & \textbf{cluster mmd} & \textbf{spectra mmd} & \textbf{graph edge overlap} & \textbf{num chambers diff} & \textbf{precision@100pagerank-hub} & \textbf{alpha gap} \\
    \toprule
    Model & \multicolumn{4}{c|}{Macro Structure} & \multicolumn{3}{c}{Macro Phenomenon} \\

    & $\mathrm{MMD.D}^2$ $\downarrow$ 
    & $\mathrm{MMD.C}^2$ $\downarrow$ & $\mathrm{MMD.S}^2$ $\downarrow$ & EO $\uparrow$ 
    & $\Delta C$ $\downarrow$  & $\mathrm{P}@100\text{-KOL}$ $\uparrow$ & $\Delta \alpha$ $\downarrow$ \\
    \midrule

    \multicolumn{8}{c}{\textbf{Propagate-En}} \\
    \midrule

    Qwen3-SFT(w/o AP) & 0.3509 & 0.4128 & 0.3739 & 0.0608 & 33 & 0.27 & 1.0884 \\
    Qwen3-RL(w/o AP) & 0.3594 & \underline{0.301} & 0.3604 & 0.0608 & 33 & 0.32 & 1.1183 \\
    Qwen3-SFT(w/ AP) & \underline{0.0526} & \textbf{0.2539} & \underline{0.2027} & \underline{0.0882} & \underline{4} & \textbf{0.37} & \underline{0.0259} \\
    Qwen3-RL(w/ AP) & \textbf{0.0351} & 0.3557 & \textbf{0.1981} & \textbf{0.1022} & \textbf{2} & \textbf{0.37} & \textbf{0.01} \\

    \midrule

    \multicolumn{8}{c}{\textbf{Weibo Tech}} \\
    \midrule
    
    Qwen3-SFT(w/o AP) & 0.2623 & 1.2628 & 0.4772 & 0.0143 & 16 & 0.3 & 1.0828  \\
    Qwen3-RL(w/o AP) & 0.2713 & 1.2345 & 0.5028 & 0.0137 & 16 & 0.28 & 1.0091  \\
    Qwen3-SFT(w/ AP) & \underline{0.1599} & \underline{1.139} & \underline{0.121} & \underline{0.0678} & \underline{9} & \underline{0.31} & \underline{0.3736}  \\
    Qwen3-RL(w/ AP) & \textbf{0.1467} & \textbf{0.7668} & \textbf{0.1027} & \textbf{0.1347} & \textbf{8} & \textbf{0.32} & \textbf{0.123}  \\

    \midrule
    \multicolumn{8}{c}{\textbf{Weibo Daily}} \\
    \midrule

    Qwen3-SFT(w/o AP) & 0.3234 & 0.8353 & 0.4558 & 0.0253 & 1 & \underline{0.44} & 1.0467  \\
    Qwen3-RL(w/o AP) & 0.6509 & 0.6874 & 0.1958 & 0.0269 & \textbf{0} & 0.27 & 1.1465  \\
    Qwen3-SFT(w/ AP) & \textbf{0.0337} & \textbf{0.4526} & \textbf{0.0314} & \underline{0.0911} & \textbf{0} & \textbf{0.49} & \textbf{0.1097}  \\
    Qwen3-RL(w/ AP) & \underline{0.0614} & \underline{0.4983} & \underline{0.0338} & \textbf{0.0973} & 1 & 0.32 & \underline{0.2074}  \\

    \bottomrule
  \end{tabular}
  \end{adjustbox}
  
  \label{tab:ablation_Graphia}
\end{table*}

\subsection{Ablation on Graphia Components}

% 我们对于Graphia的IDGG图生成过程做了abalation。我们对于两个关键的部分进行了abalation：1. 是否进行强化学习训练 2. 是否进行activity predictor的训练。
% 如Table~\ref{tab:ablation_Graphia}所示，可以观察到两个现象：1. 两部分对于图生成的效果都有正向提升；最明显的是对于edge overlap，三个数据集上两部分的训练都加上可以达到最好的效果 2. 对于相对较小的图，rl的训练效果更好；rl之后各个指标可以得到一致的提升 3. 而对于较大的图，rl的训练效果提升较为不明显，此时训练activity predictor提供degree的先验信息即可生成较好的图。由此可见，对于较大规模的图生成，需要训练一个比较好的

We conduct an ablation study to investigate the influence of two critical components in the Graphia IDGG generation pipeline:  

(1) whether reinforcement learning (RL) is applied during training; 

(2) whether the activity predictor (AP) needs to be trained to provide prior information about source node degrees.  

As shown in Table~\ref{tab:ablation_Graphia}, we identify three key findings:  

(1) Both RL and AP contribute positively to the overall quality of generated graphs. In particular, on the \textit{edge overlap} (EO) metric, the combination of RL and AP achieves the best performance consistently across all three datasets;

(2) For small-scale graphs such as Weibo-Tech and Propagate-En, RL yields more substantial improvements, with consistent gains observed across multiple evaluation metrics;

(3) For large-scale graphs, the marginal benefit of RL diminishes. In these cases, a well-trained AP that provides reliable degree priors is sufficient to generate high-quality graphs.

Overall, these results highlight that for large-scale graph generation, training an effective AP is crucial to incorporating strong structural priors, while RL training is particularly advantageous in small-scale scenarios.

\subsection{Simulation of Platform Incentives}
Through TDGG and IDGG alignment experiments, we show that the discrepancy between the Graphia-generated graph $\hat{\mathbf{G}}_{\text{fut}}$ and the reference graph $\mathbf{G}_{\text{fut}}$ remains within a controllable range. Building on this, we run counterfactual, platform interventions to test whether network shifts plausibly to incentives. We inject a single broadcast into every person's memory $\mathcal{M}_{t}(u)$: a comment-focused incentive on Weibo Daily and a repost-focused incentive on Weibo Tech. 
\begin{figure}[htbp]
  \subfloat{\includegraphics[width=\linewidth]{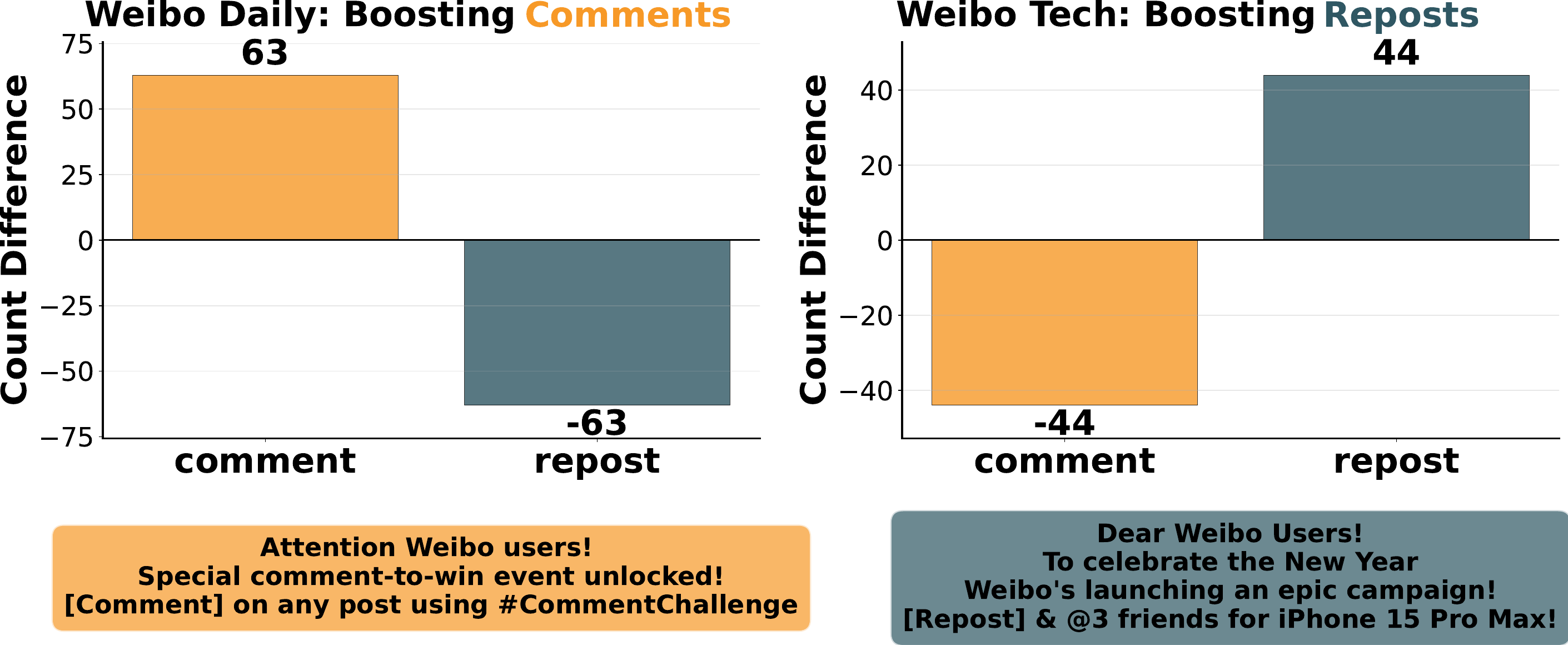
  }}
  % \vspace{1em}
\caption{Impact of broadcast incentives on message propagation in the Weibo networks.}
\label{fig:broadcast_exp}
\end{figure}

As shown in Figure~\ref{fig:broadcast_exp}, $\hat{\mathbf{G}}_{\text{fut}}$ build on Weibo Daily shift toward comments ($+63$) with a symmetric drop in reposts ($-63$), whereas $\hat{\mathbf{G}}_{\text{fut}}$ build on Weibo Tech shifts toward reposts ($+44$) with a symmetric drop in comments ($-44$). 
These results indicate that platform-level incentives can effectively steer community evolution in social graphs by reshaping the interaction patterns, which demonstrate that Graphia can support plausible counterfactual simulations.

\section{Scalability of Graphia}
\label{app:scalability}

\begin{table*}[ht]
\centering
\caption{Runtime (hours:minutes) of Graphia components across datasets. the overall training time sumed by steps for rl training, and epochs for Activity predictor training.}
\label{tab:Graphia_components}
\begin{adjustbox}{width=.85\textwidth,  keepaspectratio}
\begin{tabular}{l l c c c}
\toprule
Model & Dataset & Inference (h:m) & Train (h:m) 50 steps/epochs & Train (h:m) 100 steps/epochs\\
\midrule
Graphia-Q          & Propagate-En & 0:16 & 3:03 & 6:06\\
                   & Weibo-Tech & 0:23 & 6:46 & 13:31\\
                   & Weibo-Daily & 0:28 & 6:14 & 12:28\\
Graphia-E          & Propagate-En & 0:22 & 4:26 & 8:51\\
                   & Weibo-Tech & 0:39 & 4:17 & 8:33\\
                   & Weibo-Daily & 1:10 & 8:27 & 16:53\\
Activity Predictor & Propagate-En & 0:00 & 0:04 & 0:08\\
                   & Weibo-Tech & 0:00 & 0:05 & 0:10\\
                   & Weibo-Daily & 0:00 & 0:08 & 0:16\\
\bottomrule
\end{tabular}
\end{adjustbox}
\end{table*}

We present two aspects of scalability analysis.

\paragraph{Theoretical Analysis}
Most deep learning-based graph generative models (GGMs) capture high-order dependencies but incur super-linear time complexity, typically $O(N^2)$~\cite{ieee_survey}, which limits their applicability to small-scale graphs such as molecular structures. Only a limited number of approaches achieve linear complexity with respect to the number of edges, i.e., $O(M)$~\cite{EDGE}. Our method is inspired by EDGE~\cite{EDGE}: it first predicts node degrees using an $O(N)$ activity predictor, followed by $O(N)$ training for Graphia-Q and $O(M)$ training for Graphia-E, resulting in an overall time complexity of $O(M)$.

\paragraph{Empirical Comparison}
We benchmark the training and inference time of Graphia against GAG-General. We acknowledge that, unlike the training-free GAG-General, Graphia incurs additional training overhead.
Nevertheless, we apply several engineering optimizations, including distributed training across eight H20 GPUs using the Ray framework and asynchronous inference on four GPUs via vLLM.
The Activity Predictor component is trained separately on a single A800 GPU.
Crucially, Graphia decouples prompt processing from token generation, thereby eliminating the prompt preprocessing latency that GAG-General incurs during inference. As shown in Tables~\ref{tab:Graphia_components}, Graphia demonstrates consistently lower end-to-end generation latency across all evaluated datasets (Propagate-EN, Weibo-Tech, and Weibo-Daily). In terms of total training time, its overall cost remains comparable to that of GAG-General for large graphs like Weibo-Daily.

% \begin{table*}[ht]
% \centering
% \caption{{\revision End-to-end runtime comparison between Graphia and GAG-General.}}
% \label{tab:Graphia_vs_gag}
% \begin{adjustbox}{width=.85\textwidth,  keepaspectratio}
% \begin{tabular}{l l c c c}
% \toprule
% Model & Dataset & Inference (h:m) & Train (h:m) 50 steps/epochs & Train (h:m) 100 steps/epochs\\
% \midrule
% Graphia      & Propagate-En & 0:38 & 7:33 & 15:036 \\
%              & Weibo-Tech & 1:02 & 11:08 & 22:16 \\
%              & Weibo-Daily & 1:38 & 14:49 & 29:38 \\
% GAG-General  & Propagate-En & ---  & --- & --- \\
%              & Weibo-Tech & 4:47 & --- & --- \\
%              & Weibo-Daily & 26:39 & --- & --- \\
% \bottomrule
% \end{tabular}
% \end{adjustbox}
% \end{table*}

\section{Graph Data Construction}

In social graph simulation, we consider a sequence of time-stamped graph snapshots, $\{G_t\}_{t=1}^T$, where each $G_t = (\mathcal{V}_t, \mathcal{E}_t, \mathbf{P}_t, \mathbf{X}_t)$.  
Taking the Weibo social network as an example:
\begin{itemize}
    \item $\mathcal{V}_t$ denotes the set of Weibo users at time $t$;
    \item $\mathcal{E}_t$ denotes the set of interaction edges between users at time $t$;
    \item $\mathbf{P}_t$ represents the collection of user profile texts at time $t$;
    \item $\mathbf{X}_t$ represents the collection of interaction texts (e.g., comments or reposts) between users at time $t$.
\end{itemize}

\medskip\noindent\textbf{Example: $G_t$ at $t=3$}
\begin{verbatim}
Users: [Alice, Bob, Charlie]  
Profiles:  
    Alice: "I love tech"  
    Bob:   "Coffee lover"  
    Charlie: "Student"  
Edges:     
    Alice → Bob  
    Bob → Charlie  
Interactions: 
    Alice → Bob: "Check out this paper!"
    Bob → Charlie: "Hey, need help?"
\end{verbatim}

To encode graph context for LLM input, for a source node $u$ at time $t$, we construct the prompt as:
\begin{align*}
[p_u] + [\mathcal{M}_{t}(u)] & \rightarrow \text{LLM} \rightarrow [\text{Query}], \\
[\text{Query}] &\rightarrow \hat{C}_t^{u}, \\
[p_v, p_u, \mathcal{M}_{t}(u), \mathcal{M}_{t}(v)] & \rightarrow \text{LLM} \rightarrow [\hat{m}_{u \to v}, \hat{y}_{u\to v}].
\end{align*}
where $p_v$ is the profile of candidate destination node $v$, $\mathcal{M}_{t}(u)$ and $\mathcal{M}_{t}(v)$ are the memory banks of source node $u$ and destination node $v$ respectively, containing their historical interactions up to time $t$.
Detailed prompt templates are provided in Tables~\ref{prompt:dst_select}, \ref{prompt:edge_generation}, and~\ref{prompt:llm_as_judge}.

\section{Online Resources}
% TODO, opensource github repo and data.
Our code, data, model checkpoints, and baseline implementations are publicly available at \url{https://anonymous.4open.science/r/Graphia}.

\section{Use of Large Language Models}
LLMs are employed in two specific aspects of this work. First, we use LLMs as a writing aid to polish the manuscript text and refine figure captions, improving clarity and presentation quality. Second, our proposed framework is built upon LLM with reinforcement learning.
No other parts of the research involved significant LLM assistance.

\newpage
% \section{Prompt}

\begin{table*}[htbp]
  \caption{The template of Graphia-Q for destination selection.}
  \label{prompt:dst_select}
  \begin{tcolorbox}[colback=blue!5, % Light blue background
  colframe=black]
  You should act as a src node in the network. You are given a list of dst nodes and their node texts. 

  Your task is to predict the profile of dst nodes. You should think about the dst nodes you are going to interact with. 

  ** Objective ** 

  You should maximize the chances to retrieve desired dst nodes with your query text. 

  Your task is to depict node text of dst nodes for the src node <src id> 

  You're about to interact with <dx src> dst nodes in the network. 

  <environment description> 

  [For src-node (<src id>):] 

  <src node text> 

  Here's your interaction history with other destination nodes in the network: 

  <memory dst texts> 

  Here's your friends' interaction history with other destination nodes in the network: 

  <neighbor dst texts>
  \end{tcolorbox}
\end{table*}

\begin{table*}[htbp]
  \caption{The template of Graphia-E for edge generation.}
  \label{prompt:edge_generation}
  \begin{tcolorbox}[colback=blue!5, % Light blue background
  colframe=black]
You should generate the edge attributes for the edge (relation/action between src and dst node). 

You should think about the edge attribute.

** Objective **

You should first predict the edge LABEL. Then generate the edge TEXT, consistent with src node history edges.

[For src-node (<src id>):]

<src node text>

[For dst-node (<dst id>):]

<dst node text>

Src-node(<src id>) past edges:

<memory edge texts>
  \end{tcolorbox}
\end{table*}

\begin{table*}[htbp]
  \caption{The template of LLM-as-ajudge for edge generation evaluation. }
  \label{prompt:llm_as_judge}
  \begin{tcolorbox}[colback=blue!5, % Light blue background
  colframe=black]
You are an expert judge evaluating the quality of a response to a given prompt. Please evaluate the role-playing ability of the ACTOR NODE based on its actor actions consistency with reference actions.

The social goal for the actor is:

   <goal>

[Prompt]:

<prompt>

[ACTOR Action]:

<response>

[Reference Action]:

<reference>

Scoring Logic  

- GOAL Fulfillment(GF):

1 (Frequent mismatches with the goal),3 (Mostly aligned, with minor inconsistencies),5 (Fully aligned with the goal) 

- Contextual Fidelity(CF):  

1 (Frequent inconsistencies),3 (Minor inconsistencies),5 (Deep contextual mastery)  

- Personality Depth(PD):

1 (Contradictory traits),3 (Occasional deviations),5 (Nuanced embodiment)  

- Dynamic Adaptability(DA):  

1 (Rigid responses),3 (Context-dependent adaptation),5 (Creative innovation)  

- Immersive Quality(IQ):  

1 (Disruptive inconsistencies),3 (Minor immersion breaks),5 (Seamless portrayal)  

- Content Richness(CR):  

1 (Superficial/output),3 (Adequate detail),5 (Rich, layered interactions)

Your response must follow the format provided below/ Please note that only when the content quality is extremely good can 5 Points be given.

[Response Format]:

GF: [1-5]

CF: [1-5]  

PD: [1-5]  

DA: [1-5]  

IQ: [1-5]  

CR: [1-5]

[Response]:
  \end{tcolorbox}
\end{table*}

\end{document}